\documentclass[twocolumn]{aastex62}

\usepackage{tabularx}

\usepackage{amsmath}
\shorttitle{Modeling Dwarf-Dwarf Interactions}
\shortauthors{Pearson et al.}

\def\deg{\ifmmode^\circ\else$^\circ$\fi} 
\def\identikit{{\it Identikit}}
\newcommand{\msun}{\ensuremath{\mathrm{M}_\odot}}
\newcommand{\kms}{\ensuremath{\mathrm{km}~ \mathrm{s}^{-1}}}

\begin{document}

\title{Modeling the Baryon Cycle in Low Mass Galaxy Encounters: \\the Case of NGC 4490 \& NGC 4485}

\correspondingauthor{Sarah Pearson}
\email{spearson@astro.columbia.edu}

\author[0000-0003-0256-5446]{Sarah Pearson}
\affil{Department of Astronomy, Columbia University, New York, USA}

\author[0000-0003-3474-1125]{George C. Privon}
\affiliation{Department of Astronomy, University of Florida, Gainesville, USA}

\author{Gurtina Besla}
\affiliation{Steward Observatory, University of Arizona, Tucson, AZ, USA}

\author{Mary E. Putman}
\affiliation{Department of Astronomy, Columbia University, New York, USA}

\author{David Mart{\'{\i}}nez-Delgado}
\affiliation{Astronomisches  Rechen-Institut,  Heidelberg, Germany}

\author{Kathryn V. Johnston}
\affiliation{Department of Astronomy, Columbia University, New York, USA}

\author{R. Jay Gabany}
\affiliation{Black Bird II Observatory, Alder Springs, CA, USA}

\author{David R. Patton}
\affiliation{Department of Physics and Astronomy, Trent University, Peterborough, Canada}

\author[0000-0002-3204-1742]{Nitya Kallivayalil}
\affiliation{Department of Astronomy, University of Virginia, Charlottesville, VA, USA}



\begin{abstract}
Discoveries of low mass galaxy pairs and groups are increasing. Studies indicate that dwarf galaxy pairs are gas rich in the field and exhibit elevated star formation rates, suggestive of interactions.  Lacking are dynamical models of observed dwarf galaxy pairs to disentangle the physical processes regulating their baryon cycles. We present new optical data and the first detailed theoretical model of an observed tidal encounter between two isolated low mass galaxies, NGC 4490  \& NGC 4485. This system is an isolated analog of the Magellanic Clouds and is surrounded by a $\sim$50 kpc extended HI envelope. We use hybrid $N$-body and test-particle simulations along with a visualization interface (\identikit) to simultaneously reproduce the observed present-day morphology and kinematics. Our results demonstrate how repeated encounters between two dwarf galaxies can ``park" baryons at very large distances, without the aid of environmental effects. Our best match to the data is an 8:1 mass ratio encounter where a one-armed spiral is induced in the NGC 4490-analog, which we postulate explains the nature of diffuse starlight presented in the new optical data. We predict that the pair will fully merge in $\sim$370 Myr, but that the extended tidal features will continue to evolve and return to the merged remnant over $\sim$5 Gyr. This pre-processing of baryons will affect the efficiency of gas stripping if such dwarf pairs are accreted by a massive host. In contrast, in isolated environments this study demonstrates how dwarf-dwarf interactions can create a long-lived supply of gas to the merger remnant.
\end{abstract}


\section{Introduction} \label{sec:intro}
The impact of mergers on the structure and gas content of massive galaxies has been studied extensively both theoretically  (e.g. \citealt{toomre72}, \citealt{barnes88}, \citealt{springel99}, \citealt{dubinski99}, \citealt{barnes16}) and observationally (e.g. \citealt{arp66},
\citealt{sanders88}, \citealt{engel10}, \citealt{bussmann12}
). However, the merger sequence and any consequent morphological transformation of low mass galaxies (M$_* < 10^{10}$ M$_\odot$) through tidal processes is not well constrained. There is reason to believe that the merger sequence of dwarf galaxies could differ substantially from that of massive galaxies.  Dwarfs in the field have higher gas fractions (\citealt{geha12}, \citealt{bradford15}), higher dark matter to baryon ratios (e.g. \citealt{tolstoy09}) and dwarf mergers are more numerous per unit volume than massive galaxy mergers (\citealt{fak10}).  

In this work we present the first detailed model of an observed isolated low mass galaxy pair, NGC 4490/4485.  We use this model to study in detail the role of tidal encounters with companions in the morphological evolution of low mass galaxies.  Detailed dynamically matched models to real systems are needed to age-date the systems, constrain the initial encounter parameters and to understand the timescales involved in the gas cycling due to the interactions.  While generic dwarf-dwarf mergers have been modeled in the literature (e.g. \citealt{bekki08}, \citealt{kim09}), only one observed dwarf-dwarf interaction has been modeled in detail, namely the Large and Small Magellanic Clouds (LMC and SMC) (e.g. \citealt{gardiner96}, \citealt{bekki05}, \citealt{conners06}, \citealt{besla10}, \citealt{besla12}, \citealt{diaz11},\citealt{guglielmo14}, \citealt{pardy18}). Without additional models of isolated, interacting dwarf systems, we cannot assess whether the LMC/SMC scenario is typical of dwarf interactions. 
 
Observational studies of dwarf pairs and groups are growing (\citealt{tully06}, \citealt{stierwalt17})  providing insight to the role of dwarf-dwarf interactions in the evolution of low mass galaxies. The TiNy Titans Survey (TNT: \citealt{stierwalt15}) showed that dwarf galaxy pairs in the field ($>$ 1.5 Mpc from a massive galaxy) appear just as gas rich as their non-paired counterparts, despite exhibiting elevated star formation rates relative to unpaired field dwarfs. Recent work by \citet{privon17} suggests that dwarf interactions trigger large-scale interstellar medium (ISM) compression, rather than nuclear starbursts often associated with massive mergers.  The high gas fractions of dwarfs in the field indicate that dwarfs (with M$_* > 10^7$ \msun) do not fully exhaust their gas through tidal interactions or internal processes (star formation, feedback etc.,  e.g. \citealt{bradford15}) even with their shallower potential wells (e.g. \citealt{lelli14}). 
However, the TNT and  \citet{bradford15} works are based on single dish neutral hydrogen observations (HI). Without resolved imaging it is unclear if the gas is still located within the galaxies or is spatially extended. 

This motivated the Local Volume TiNy Titans Survey (LV-TNT: \citealt{pearson16}), where we investigated resolved synthesis  HI maps and surface density profiles for 10 dwarf galaxy pairs located within 25 Mpc of the Milky Way (MW). 
We found that tidal interactions between low mass galaxies can ``park"  gas at large distances and that the gas is only prevented from being re-accreted to the dwarfs if the pairs are in the vicinity of a massive galaxy, as in the case for the Magellanic System. The gas at large distances is not actively participating in the formation of stars, which helps explain why the single dish TNT survey found that dwarf pairs with elevated SFRs could still have high gas fractions. The low mass galaxy pair NGC 4490/4485 is an example of one of the LV-TNT pairs with gas at large distances and will be the focus of this paper. This system is an isolated analog of the Magellanic Clouds that is surrounded by a massive, spatially extended HI complex ($> 50$ kpc in extent \citealt{clemens98}). 

In particular, we present a detailed $N$-body simulation that simultaneously reproduces the observed present-day morphology and kinematics of NGC 4490/4485 using  \identikit ~(\citealt{barnes09}). \identikit~ is a hybrid $N$-body and test-particle simulation, which enables a rapid exploration of the parameter space of galaxy mergers. 
The goal of this study is to utilize \identikit~ to test whether interactions between NGC 4490/4485 can naturally explain the origin of the observed extended gas complex surrounding the galaxy pair or whether other mechanisms, such as outflows, are necessary (e.g. as suggested by \citealt{clemens98}). Additionally, we aim to investigate the timescales involved in cycling gas in an isolated dwarf galaxy interaction and consider the affect of the interaction on the dwarfs involved. 

\begin{figure*}
\centerline{\includegraphics[width=\textwidth]{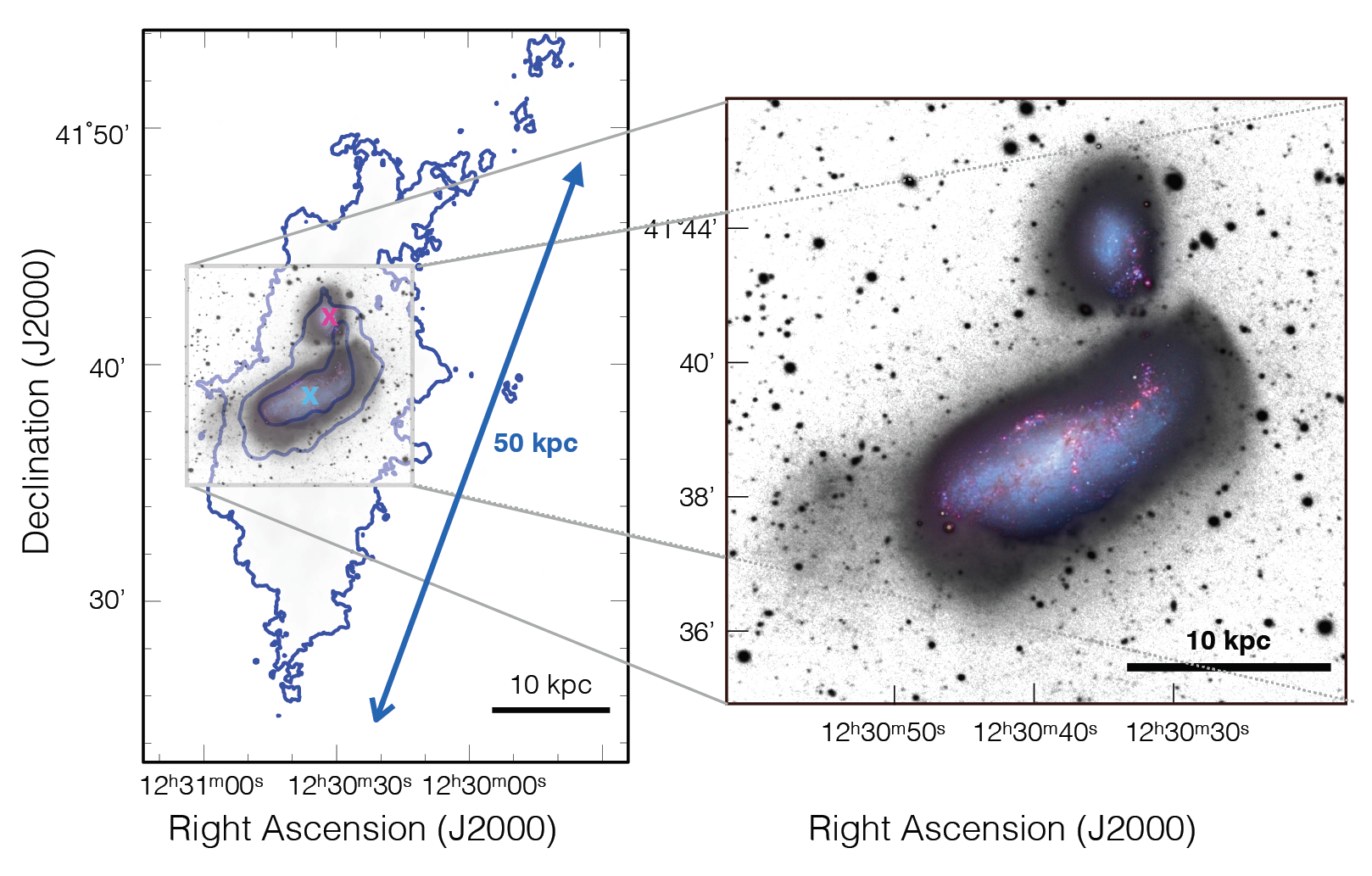}}
\caption{{\bf Left:} Neutral hydrogen (HI) envelope surrounding the dwarf galaxy pair NGC 4490 and NGC 4485 (blue: N(HI) = 0.7, 7, 35 $ \times 10^{20}$ atoms cm$^{-2}$). See \citet{clemens98}, figure 1 for additional and lower column density contours. The x's correspond to the optical centers of NGC 4490 (cyan) and NGC 4485 (magenta). The 10 kpc and 50 kpc scale bars are plotted assuming a distance of 7.14 Mpc (\citealt{theureau07}) to the dwarf galaxy pair. The HI is distributed roughly symmetrically around the dwarf pair (outer contour) and an HI bridge extends from the more massive NGC 4490 dwarf towards the smaller companion (inner contour). The gray box shows the optical data. {\bf Right:} Optical deep image of NGC 4490 and NGC 4485 obtained with the BBRO2 0.5-meter telescope (see Section \ref{sec:optical}). The surface brightness limit of this image is $\sim$ 29 mag arcsec$^{-2}$, revealing a very faint plume of stars on the East (left) side of the NGC 4490's main body. A color inset of the disk of the galaxies taken with the same telescope is included for reference.}
\label{fig:n4490}
\end{figure*}

If dwarf-dwarf tidal interactions are shown to be capable of ``parking" gas at large distances, there will be important implications to our understanding of the baryon cycle of low mass galaxies. Specifically, hierarchical processes can enable a long-lived gas supply channel for future star formation. Additionally, if the gas remains extended for a long period of time following the interaction, this could greatly affect the efficiency at which gas is stripped from these systems if they fall into a gas rich environment, such as a galaxy cluster or the CGM of a massive galaxy. 
We follow up the best dynamical match with self-consistent $N$-body simulations that test the match.  We do not include hydrodynamics in this study, as we present here a first step towards addressing the plausibility that tidal interactions between a low mass galaxy encounter can generate tidal debris to large distances. Hydrodynamics should not strongly affect the large scale tidal features (\citealt{barnes96}). 

We will utilize these simulations to explicitly define the NGC 4490/4485 system's current dynamical state, encounter history and future fate. We further compare the resulting $N$-body simulation to new optical data of the system from the f/8.1 Ritchey-Chretien 0.5-meter telescope of the Black Bird Observatory 2 (BBRO2) to investigate the consequences of dwarf-dwarf tidal interactions to the internal stellar morphology of the galaxies. 

The paper is organized as follows: in Section \ref{sec:obs} we present the dwarf pair, NGC 4490/4485. In Section \ref{sec:models} we describe our dwarf models and the matching process using \identikit~ (\citealt{barnes09}). In Section \ref{sec:results} we show the results of our dynamical match to the NGC 4490/4485 pair. We discuss the fate of the HI envelope in Section \ref{sec:disc}. In Section \ref{sec:disc2} we compare our results to the Magellanic System and we discuss the implications of tidal pre-processing and the inflow of gas to the merger remnant. We conclude in Section \ref{sec:con}.

\section{THE NGC 4490/4485 PAIR}\label{sec:obs}

The low mass galaxy pair NGC 4490/4485 (presented in Figure \ref{fig:n4490} and with optical centers marked by cyan and magenta ``x"s, respectively)  is a slightly more massive analog of the LMC/SMC, with a stellar mass ratio of $\sim$ 8:1 (\citealt{clemens98}) and isolated from any massive galaxy. The NGC 4490/4485 galaxies are separated by only 7.5 kpc in projection (the projected separation of the LMC/SMC is 11 kpc). See Table \ref{tab:prop} for a comparison of the two dwarf pairs. NGC 4490/4485 resides  7.14 Mpc from the Milky Way (\citealt{theureau07}). The nearest massive galaxy (defined as  
 M$_* > 10^{10}$ \msun~ as in \citealt{geha12}) is NGC 4369 which has a stellar mass of M$_* = 2.6 \times 10^{10}$ \msun. The projected separation between the pair and NGC 4369 is $d_{\text{proj}}>$ 300 kpc and the velocity separation to the pair is $v_{\text{sep}} >$ 400 ~\kms~ (\citealt{pearson16}). Cosmologically, dwarf galaxy pairs do not remain bound to each other for long when in proximity to a massive galaxy (\citealt{gonzalez16}). The relatively isolated environment of the NGC 4490/4485 system thus allows us to examine the evolution of a low mass galaxy pair, independent of environmental factors (see \citealt{theis01} and \citealt{paudel17} for examples of simulated dwarf interactions matched to morphological data).

\begin{table*}
\centering
\caption{Properties of NGC 4490/85 and the LMC/SMC} 
\label{tab:prop}
\begin{tabular}{lccc}
\hline
 &{\bf NGC 4490/4485} & {\bf LMC/SMC} &{unit}\\
\hline
Average distance  &  7.14\footnote{Average distance to pair from \citet{theureau07}. $^b$Average distance to pair from \citet{cioni00}. $^c$Stellar mass from \citet{marel02}. $^d$Stellar mass from \citet{stani04}. $^e$HI mass within 2MASS extent: \citet{pearson16}. $^f$HI mass residing beyond the 2MASS extents of the galaxies: \citet{pearson16}. $^g$Calculated from the velocity of the gas associated with the optical centers of both galaxies: \citet{clemens98}. $^h$As estimated from their V$_{los}$: \citet{besla12}, \citet{kalli13}.} &0.055$^{b}$ & Mpc\\
Stellar mass &7.2/0.82& 2.7$^{c}$/0.3$^{d}$ & $\times 10^9$ \msun\\
HI mass$^{e}$ & 2.4/0.23&  0.27/0.26  & $\times 10^9$ \msun\\
HI mass outside$^{f}$ &1.07& 0.37  & $\times 10^9$ \msun\\
Projected separation & 7.5 & 11  &kpc \\
Velocity separation &30$^{g}$ & 116$^{h}$   &  \kms\\
Massive host galaxy & NGC 4369 & Milky Way & \\
Av. dist. to host& 310&  55& kpc\\
\hline 
\end{tabular}
\end{table*}

\subsection{Archival HI data}\label{sec:hi}
The system is clearly detected in HI and an HI envelope, first discovered by \citet{huch80}, symmetrically surrounds the NGC 4490/4485 pair and extends $ \sim$50 kpc in projection.  

The HI data presented in this paper are originally from \citet{clemens98} and were obtained with the VLA in C-configuration and D-configuration. The velocity of the pair's HI ranges from $-123 $ ~\kms~ to 83 \kms~ centered at a systemic heliocentric radial velocity of $v_{sys} = $ 575 \kms, and the data have a velocity width per channel of 20.7 \kms. The envelope is detected to column densities of N(HI)$ \sim 3 \times 10^{19}$ atoms cm$^{-2}$, which is close to the sensitivity limit of the data (N(HI) $\sim 10^{19}$ atoms cm$^{-2}$: \citealt{clemens98}). The total gas mass of the system is $M_{\rm HI}\sim 3.7 \times 10^9$ \msun, where $\sim$ 30\% of the gas ($\sim 1.07 \times 10^9$ \msun, see Table \ref{tab:prop}) resides beyond the stellar extents of the two galaxies (defined as beyond the 2MASS extents of each disk; \citealt{pearson16}). Based on the relatively isolated environment of the pair, \citet{pearson16} found it unlikely that ram pressure is playing a significant role in the origin of the extended envelope surrounding the pair. 

There is a dense bridge (N(HI)$> 3.5 \times 10^{21}$ atoms cm$^{-2}$) of gas connecting the pair (see Figure \ref{fig:n4490}, left: inner HI contour) which suggests a tidal encounter between the galaxies (\citealt{toomre72}). The bridge material peaks at $v\sim -123$ to $-82$ \kms~ with respect to the systemic velocity. GALEX UV data of the system (\citealt{smith10})  show that stars are forming in the NGC 4490/4485 bridge. This suggests that the high gas density in the bridge is not due to a chance projection of overlapping gas. 

Interestingly, \citet{lee09} found that NGC 4485 is likely undergoing a starburst, as its H$\alpha$ equivalent width is EW $= 76 \pm 13$, which exceeds the logarithmic mean by 2$\sigma$ limit when compared to the $\sim$ 300 dwarfs in the 11HUGS Dwarf Galaxy Survey (\citealt{lee09}). NGC 4485's star formation rate inferred from the far ultraviolet (FUV) non-ionizing continuum is SFR(FUV) = 0.22 $\msun$ yr$^{-1}$ (assuming a distance of 7.14 Mpc; see \citealt{lee09} table 1). This is consistent with findings of the TNT survey that the secondary (smaller) galaxy in a dwarf galaxy pair is more likely to be starbursting, and with theoretical  expectations of stronger tides acting on the secondary. When compared to the \citet{lee09} sample, both NGC 4490 and NGC 4485 lie above the 11HUGS $M_B$ vs log(SFR) mean, but are within the scatter of the sample (see \citealt{pearson16}).

Using HI data presented in \citet{viallefond80}, \citet{elm98} computed a rotation curve for NGC 4490 and found that it peaks at a radius of 6 kpc from the center with $v_{rot}$/sin($i$) $\sim$ 80 \kms, where $i$ is the inclination of the primary in the sky plane, which is $\sim60 \deg$. A peak rotational velocity of $v_{\rm rot,peak} = 80 ~\kms$, is similar to that of the LMC ($v_{\rm rot,peak} \sim 90 ~\kms$: \citealt{marel14}). These findings are consistent with the \citet{clemens98} HI data (see their figure 2). 

Kinematic constraints are important in breaking model degeneracies that arise when matching only the morphology of galaxy mergers (e.g. \citealt{barnes09}). In this work, we assume that the optical centers are approximately the dynamical centers of the two galaxies. 
Exploring the HI data cube, we find that the gas associated with the optical center of NGC 4485 is at  $v\sim-20$ \kms~ with respect to the systemic velocity. The optical center of NGC 4485  is located to the left of the bridge material (see magenta x vs blue inner HI counter in Figure \ref{fig:n4490}, left). 
Hence, the position of NGC 4485 is offset from the bridge in position and kinematics, and has lower column densities. 
\citet{clemens98} found the gas associated with the optical center of NGC 4490 to be at $+$10 \kms~ with respect to the systemic velocity. As the velocity channel widths are 20.7 \kms, we use a velocity center between 0 \kms~ and 20 \kms~ with respect to the centered HI cube as our constraint for the line-of-sight velocity.

\subsection{Deep optical imaging}\label{sec:optical}
Optical data provides further support of a tidal encounter between NGC 4490/4485 (\citealt{elm98}, figure 3). In Figure \ref{fig:n4490} , right, we present new optical data of the NGC 4485/4490 system obtained with the f/8.1 Ritchey-Chretien 0.5-meter telescope of the Black Bird Observatory 2 (BBRO2) in Alder Springs (California) during different dark-sky nights between April 8th-22th, 2012. A 16 mega-pixel Apogee Imaging Systems U16M CCD camera was used, with $31.3 \times 31.3$ arcmin field of view and a 0.46 arcsec pixel scale. We acquired a total of 22.66 hours of imaging data in 46 half-hour sub-exposures, using a non-infrared clear luminance Astrodon E-series filter (e.g. see Figure 1 in \citealt{delgado15}). Each sub-exposure was reduced following standard procedures for dark-substraction, bias-correction and flat-fielding (\citealt{delgado09}). The surface brightness limit of this image is $\sim$ 29 mag arcsec$^{-2}$ (see the method for obtaining the surface brightness limit in \citealt{delgado10}). We see evidence of a tidal encounter through a faint stellar extension to the left of the NGC 4490 galaxy's main body (Figure \ref{fig:n4490}, right), which interestingly was also identified in \citet{elm98} in the B band at a surface brightness of 23.5 mag arcsec$^{-2}$ and in  SDSS gri imaging\footnote{\url{http://ned.ipac.caltech.edu/img/2011A+A...532A..74B/gri/PGC_041333:I:gri:bbl2011.jpg}} (\citealt{baillard11}) which has a typical surface brightness limit of 24.5 mag/arcsec$^2$. In \citet{elm98} and SDSS the extension is detected as a narrower tail-like feature extending from the northern part of the NGC 4490 main body. We confirm the existence of this structure and our deeper data reveal this structure to be spanning larger distances ($\sim$ 8 kpc) in a more plume-like feature extending from the plane of the NGC 4490 disk. There is also HI gas at the location of the plume-like feature (see \citealt{clemens98} for additional HI contour levels, where N(HI)$_{\rm min}$ = 2.34 $ \times 10^{19}$ atoms cm$^{-2}$).

To summarize, the existence of a bridge connecting the two galaxies and a starburst in the secondary is suggestive of strong tidal interactions between the two galaxies that may have resulted in the observed extended gaseous envelope. The NGC 4490/4485 pair is therefore an excellent candidate for testing a dynamically-driven formation scenario for an extended HI envelope through a dwarf-dwarf encounter. Despite such tidal encounters, the primary NGC 4490 still possesses a disk with a well-defined rotation curve.  We seek to reproduce these properties using a tidal interaction model.  

\section{Dynamical Simulations}\label{sec:models}
When modeling the mergers of galaxies, several degrees of freedom exist and searching the full parameter space of a galaxy merger can therefore be time consuming. As discussed in detail in \citet{barnes09}, to model the dynamical interaction of two disk galaxies, the mass ratio ($\mu$), disk orientations ($i_1$, $\omega_1$) and ($i_2$, $\omega_2$) following the definition in  \citet{toomre72}, the eccentricity of the orbit ($e$) and the pericentric separation ($r_{peri}$) need to be specified. Hence, seven parameters are needed to model the dynamical interaction, without accounting for the internal structure of the galaxies. An additional nine parameters are needed to compare the models to observations: the length scale ($L$), the velocity scale ($V$), the center of mass position on the plane of the sky ($X_m, Y_m$), the center of mass velocity ($V_c$), the viewing angles ($\theta_x, \theta_y, \theta_z$) and the time of viewing ($t$). Without varying the internal structure of the galaxies involved (e.g. dark matter to baryon fractions, scale lengths etc.) 16 free parameters are present. 
 
In this work, we use \identikit~ (\citealt{barnes09} and   {\it Zeno} (\citealt{barnes11}) 
to  explore the parameter space of the NGC 4490/4485 interaction. Our procedure is summarized as follows:
\begin{itemize}
\item[1.]{We build a library of encounters using hybrid test-particle disks embedded in live $N$-body dark matter halos with a fixed orbital eccentricity, and vary the mass ratios and pericentric separations of the two galaxies.} 
\item[2.]{We load the test-particle simulations and projections of the data into the \identikit~ visualization interface and vary the disk orientations, viewing angles, time of viewing, scaling and center of mass position of the galaxies in order to identify the best match to the HI kinematic and morphological data of the NGC 4490/4485 system.}
\item[3.]{Based on our best match obtained with the test-particle simulations, we run a small set of $N$-body simulations of galaxy encounters with self-gravitating disks embedded in live dark matter halos. For these we fix the eccentricity, mass ratios, disk inclinations and scalings, but allow for the time and viewing angles to be varied along with a small set of pericentric separations.}
\end{itemize}

\identikit~ was initially tested by \citet{barnes09} on 36 artificially constructed mergers of massive galaxies. 
They demonstrated that, in cases where the merging system displayed prominent tidal features, the viewing directions, spin orientations and time since pericenter were well recovered, while the pericenter separation showed the largest scatter. Additionally, the velocity scalings showed a $\sim10\%$ bias for dynamically cold massive galaxy disks, as the test-particles in \identikit~ have zero velocity dispersion. 

Re-simulating the galaxy interactions with a self-gravitating disk is an important test to verify the overall morphology. The global morphology and kinematics (the focus of this paper) should not be strongly modified by self-gravity in contrast to self-gravitating features such as spiral arms (\citealt{privon13}). See \citet{barnes09} for a detailed description on the \identikit~ methodology and visualization techniques. 

In the following, we describe the details of the galaxy mass models (Section \ref{sec:massmodel}), the library of \identikit~ test-particle simulations and the matching procedure (Section \ref{sec:test}) as well as the self-gravitating $N$-body follow-up simulations  (Section \ref{sec:nbody}).

\subsection{Galaxy Models}\label{sec:massmodel}
Using {\it Zeno} (e.g. \citealt{barnes09}, \citealt{barnes11}), we set up two galaxy mass models mimicking the primary (more massive) dwarf (NGC 4490) and secondary (less massive) dwarf (NGC 4485), respectively.  We construct galaxy mass models with the same parameters as presented in \citet{barnes09} but we omit the bulge, such that our galaxy models are more dwarf-like. See \citet{barnes09} Appendix B for a detailed description of the galaxy construction in {\it Zeno}. The galaxy models were set up in approximate initial dynamical equilibrium (\citealt{barnes12}). 
Each galaxy consists of an NFW dark matter halo (\citealt{navarro96}), which tapers at large radii ($b_{halo}$) following \citet{springel99}: 

\begin{equation}
\rho_{halo} (r) = \frac{m_{halo}(a_{halo})}{4\pi ({\rm ln(2)-1/2)}}\frac{1}{r(r+a_{halo})^2}, r\leq b_{halo}\\
\end{equation}
\begin{equation*}
 = \rho^*_{halo} \left(\frac{b_{halo}}{r}\right)^{\beta} e^{-r/a_{halo}}, r > b_{halo}
\end{equation*}
where $m_{halo}(a_{halo})$ is the halo mass within the scale radius of the halo ($a_{halo}$) and $\rho^*_{halo}$ and $\beta$ are fixed by requiring that both $\rho_{halo}(r)$ and $d\rho_{halo}/dr$ are continuous at $r = b_{halo}$. 

The disk follows an exponential radial profile (\citealt{freeman70}) and a sech$^2$ vertical profile (\citealt{vander81}):

\begin{equation}
\rho_{disk} (q,\phi,z) =\frac{m_{disk}}{4\pi r_s^2z_{disk}} e^{-q/r_s} {\rm sech}^2(z/z_{disk}),
\end{equation}
where ($q = \sqrt{x^2 + y^2}, \phi, z)$ are cylindrical coordinates, $r_s$ is the disk scale radius, $z_{disk}$ is the disk scale height and $m_{disk}$ is the mass of the disk. For the secondary galaxy, we setup mass models with 1/4th, 1/6th and 1/8th of the mass of the primary galaxy. The scale length of the smaller galaxy is scaled accordingly to maintain constant mass surface density. 

In both galaxies, 20\% of the mass is made up by baryons and 80\% of the mass is dark matter (see Section \ref{sec:prop} for the affect of variations on these values). The ratio of the disk scale radius, $r_s$, to dark matter halo scale radius, $a_{halo}$, is ${r_s}/{a_{halo}} ={1}/{3}$ in both the primary (more massive) and secondary (less massive) galaxy model. The galaxy encounters are run in simulation units and then the scale factors are fit during the matching process.
All model parameters are listed in Table \ref{tab:model}. 

We ran isolated realizations of both galaxy mass models to check their long-term stability. Initially both galaxies are unstable to bar formation due to omission of the bulge.
A bar forms after 2 rotation periods at 3 disk scale lengths ($t_{\rm sim}$ = 2, see Table \ref{tab:model} for the scale from simulation to physical units in our match)\footnote{Note, we ran the galaxies in isolation and re-ran a simulation of the encounter using disks in which the bars had formed and settled prior to the encounter. This did not change the overall results of our preferred match. However, the specific location of induced spiral arms might be affected slightly by the bar's phase at the time of match (see also \citealt{barnes04}, \citealt{privon13} for a discussion on bar misalignments in the Mice: NGC 4676A/B).}.

\subsection{{\it Identikit} test-particle simulations and matching}\label{sec:test}
To explore the parameter space of the NGC 4490/4485 encounter, we build a library of dwarf encounters with various mass ratios ($\mu = 4:1, 6:1, 8:1$) and initial pericentric separations ranging from $r_{peri} = [0.75 - 5.25] \times r_{s,prim}$ in increments of 0.75 $\times r_{s,prim}$, where $r_{s,prim}$ is the disk scale length of the primary galaxy. We limit our investigation to $e=1$ orbits as e $<$ 1 orbits imply a previous encounter (see \citealt{barnes09}). The eccentricity will evolve over the course of the encounter owing to dynamical friction. 

We create two self-consistent galaxy halos (see previous section) and specify the eccentricity of the orbit and the pericentric separation. We place the galaxies' initial positions such that their dark matter halos are not overlapping and the velocities are set based on an initial idealized point-source Keplerian orbit. The galaxies are evolved in time using a standard treecode (\citealt{barneshut86}, \citealt{barnes11}) and initially follow a Keplerian orbit, however the orbit rapidly decays due to the dynamical friction from the live $N$-body dark matter halos.  The initial pericentric pass of the idealized point-source Keplerian orbit is slightly smaller (see Section \ref{sec:results}) than the initial pericentric pass including the extended live $N$-body dark matter halos. Throughout the paper we quote the non-idealized pericentric separations. 

Both dark matter halos are populated with a spherical distribution of massive particles that has the same cumulative radial mass distribution as for the initial stellar disks. These spheres are populated with multiple disks of test particles on circular orbits. This is similar to populating each galaxy with all possible disk configurations, although the test-particles do not have mass and are therefore not self-gravitating (see \citealt{barnes09} for a detailed description). 

After completing a library of test-particles simulations, we load them into the \identikit~ visualization software along with the HI data (see Section \ref{sec:morph}) and decide which disk to display. 
Subsequently, we require the simulated galaxies' positions  ($X_m$, $Y_m$) and velocities ($V_c$) to agree with the observed galaxy positions and velocities. In real time, we then vary the disk orientations ($i_{prim}$, $\omega_{prim}$, $i_{sec}$, $\omega_{sec}$), the viewing angles ($\theta_x, \theta_y, \theta_z$) and the scaling ($L,V$) of the system,  while stepping through time ($t$), mass ratio ($\mu$) and pericentric separations ($r_p$) until we obtain a good match to the data (see \citealt{barnes09}, \citealt{privon13} for a detailed description of the matching procedure).  

The key features we aim to reproduce are: 1) the position of the two galaxies; 2) the tidal debris populating the symmetric envelope (north and south) morphologically and kinematically; 3) the optical centers of NGC 4490/4485 in position (see x's in Figure \ref{fig:n4490}, left) and velocity space; 4) that the orbit of the secondary passes through the dense bridge material (see Figure \ref{fig:n4490}, left, inner blue contour) both in the morphological and kinematic panels of the \identikit~ visualization interface in order to have a plausible formation scenario for the dense bridge (e.g. ram-pressure from passing through the NGC 4490 disk:  \citealt{condon93},  \citealt{clemens00}, \citealt{gao03}). We also test matches in which the orbit does not pass through the dense bridge material in the data (see {\it Appendix}), as the bridge could in principle be purely tidal (\citealt{toomre72}).   

Using the test-particle simulations visualized in the \identikit~ interface, we scale the system so that it matches the observed extent on the sky. The scale lengths of the galaxies might be affected by the tidal interaction, so the scale length inferred for the progenitor galaxy does not necessarily have to be the present-day scale factor. However, once we find a good match we use the scale lengths as a sanity check such that we obtain a physical size scale mimicking these types of galaxies. In particular, we require that the scale length of the primary disk, $r_{s,prim}$, is at least $0.7$ kpc. None of our matches had scale lengths larger than that of the LMC disk scale length in the \citet{besla12} models (1.7 kpc), hence we did not set a strict upper limit for the scale length when exploring our matches. Once we obtain a physical scaling, we search for test-particle simulation matches that approximately reproduce the physical separation of the galaxies at present day (i.e. $\sim 7\pm2$ kpc in projection). In addition to the physical size scale, we ensure that the extent of kinematic data mimics that of our primary galaxy as we match our simulation output to the HI data in the \identikit~ visualization interface.  As a sanity check, we also compute the rotational velocity curve for our primary galaxy and check whether it is consistent with the observed HI rotational velocity curve from \citet{elm98}, peaking at $v_{\rm rot,peak} \sim 80 ~\kms$.

\subsection{Self-consistent N-body simulations}\label{sec:nbody}
After obtaining a dynamical match to the system using the test-particle simulations (see Section \ref{sec:test}, \ref{sec:match}), we  run a self-consistent collisionless $N$-body simulation including self-gravity of the disks. We compare this simulation to the test-particle simulation to check whether a more realistic disk treatment changes our match to the tidal features.  
Using the $N$-body follow-up with self-gravitating disks, we investigate if the match is affected by the slightly different dynamical friction that arises due to the higher local density in the self-consistent stellar disks.  
We do not include hydrodynamics in our simulations and the baryonic mass is therefore assumed to be a combination of stars and gas throughout this paper.  These simulations will be followed up with full hydrodynamics in future studies. The goal of this study is to assess the plausibility that tides can create tidal structures similar in extent and kinematics as the observed HI envelope. 

The self-gravitating $N$-body model presented in this paper utilizes the mass models in Section \ref{sec:massmodel} and encounter parameters based on the best match using the test-particle simulations, introduced in Section \ref{sec:match}. To test the sensitivity of our best-match to encounter parameters in the full $N$-body simulation, we also run three different initial pericentric separations, $r_p$, centered on the value obtained for the best match in the test-particle simulations. 

We load the output of the self-gravitating encounter into the \identikit~ visualization software with the same viewing angles, ($\theta_x, \theta_y, \theta_z$), and scalings ($L$, $V$) as obtained for our best match in our test-particle simulations.  Subsequently, to investigate potential differences between the test-particle match and the  $N$-body follow-up, we test the viewing angles, scaling of the system and pericentric separation at different points in time in the simulation to assess the quality of the match to the observational data.

\section{Results}\label{sec:results}
In this section we detail the best-match parameters (see Section \ref{sec:test}) obtained using the \identikit~ test-particle simulations and visualization interface and we present and analyze the $N$-body follow-up with self-gravitating disks (Section \ref{sec:match}).
Additionally, we describe the formation mechanism of the extended tidal envelope (Section \ref{sec:baryon}) and the morphological consequences of the interaction on the primary galaxy (Section \ref{sec:prim}).
We reiterate that the  goal  of  our  study    is  not  to  reproduce every detail of the NGC 4490/4485 system, but to explore whether there is a plausible dynamical solution for which the kinematics and morphology of its baryonic distribution can be quantitatively matched through a tidal encounter between the two galaxies. This is the first time \identikit~ has been used to simulate a dwarf-dwarf merger. 

\begin{figure*}
\centerline{\includegraphics[width=\textwidth]{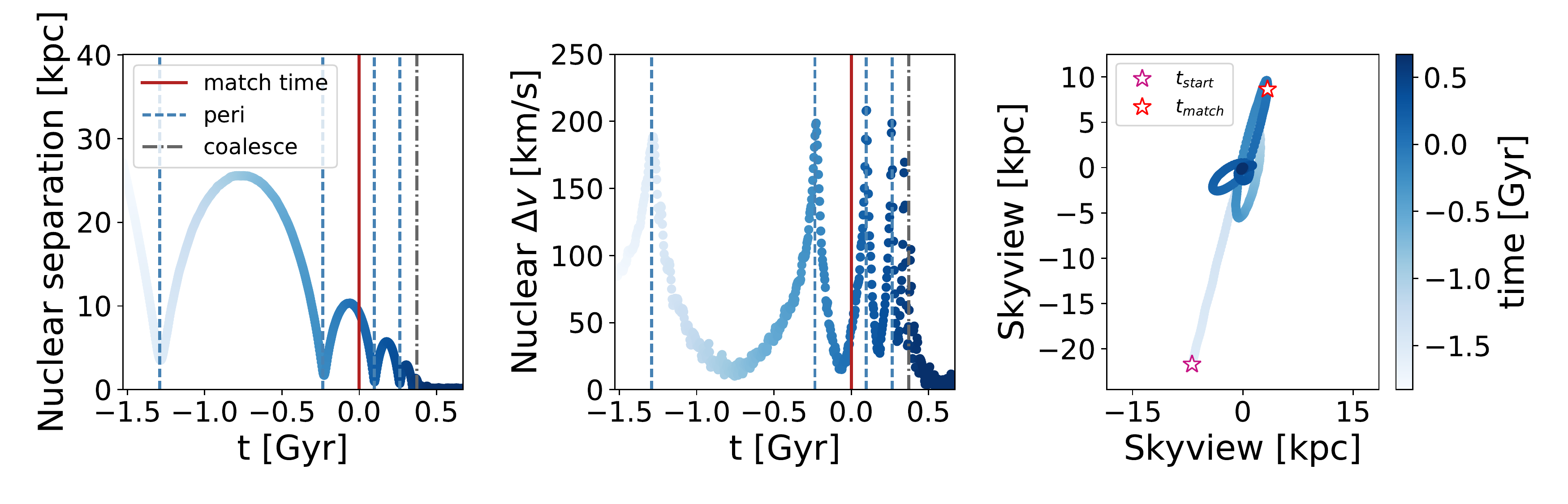}}
\caption{{\bf Left:} Nuclear separation of the two galaxies as a function of time demonstrating the secondary's (NGC 4485 analog)  orbital decay. The initial orbit is a parabolic ($e = 1$) orbit with a mass ratio of 8:1 between the galaxies and a first pericentric separation of 3.5 kpc. The physical scaling is based on the best match (see Table \ref{tab:model}) and assuming a distance of 7.14 Mpc (\citealt{theureau07}) to the system. The color gradient encodes the evolution in time relative to the start of the simulation (white: simulation start time). The dashed blue lines show the time of each pericentric passage while the red solid line shows the time of the best match to the system ($t = 0$ Gyr). The match is located near the second apocenter and we predict the pair will coalesce $\sim$ 370 Myr after the time of match (gray vertical line). At the time of match the 3D nuclear separation is $\sim 9.3$ kpc. {\bf Middle:} Nuclear velocity separation of the two galaxies as a function of time since the start of the simulation. At the time of match the velocity separation ($\Delta v$) is ($\sim 43$ \kms~). 
{\bf Right:} Orbit of the secondary about the primary, projected onto the plane of the sky for our best-fit viewing directions. The magenta star indicates the initial position of the secondary galaxy in the simulation, and the red star indicates the position of the secondary at the time of match. The orbit is centered on the position of the primary. As this is a high mass ratio merger (8:1) the center of the primary galaxy only moves slightly in response to the secondary's orbital evolution. The fact that we observe the system today (and that the two galaxies have not yet fully merged) indicates that they formed very far apart and that it is unlikely for the pair to have survived as a binary for a Hubble time.}
\label{fig:orbit}
\end{figure*}

\begin{table*}
\centering
\caption{Self-consistent $N$-body run of best match}
\label{tab:model}
\begin{tabular}{lrrr}

\hline 
{\bf \large  {\it Identikit} match scalings} &simulation units (sim) &physical \\
\hline
Velocity 	 & 1&82.5 km s$^{-1}$  \\
Time  & 1&118.7 Myr  \\
Length 	 & 1&10.02 kpc \\
Mass	  & 1&1.58 $\times 10^{10}$ M$_{\odot}$ \\
&&\\
\hline
& {\bf Primary Galaxy }& {\bf Secondary Galaxy} \\ 
\hline
{\bf \large Simulation properties} 	 & &\\
\hline
Grav. soft. (sim)/physical	 	&(0.00375)/0.0375 kpc &\\
Particle no. dark matter		&65536&32768	\\
Particle no. baryonic	&65536&32768\\
Particle mass dark matter		&244,141 M$_{\odot} $&61,035 M$_{\odot} $	\\
Particle mass baryons	&61,033 M$_{\odot} $&15,259  M$_{\odot} $\\
&&\\
\hline
{\bf \large Dynamical properties} 	 & &\\
\hline
{\bf \large t$_{start}$} (initial conditions)	 & &\\
\hline
$m_{\rm halo}$	 (sim)/physical	     &(1)/$1.6  \times 10^{10}$ M$_{\odot}$	& $(0.125)/2  \times 10^9$ M$_{\odot}$\\
$m_{\rm disk, baryons} $(sim)/physical	  	 & (0.25)/4 $\times 10^9$ M$_{\odot} $&(0.03125)/0.5 $\times 10^9$ M$_{\odot} $ 	\\
f$_{\text{\rm baryon}}$ 		&0.25 & 0.25\\
$r_{p,1}$ (idealized Keplerian orbit) & (0.25)/2.5 kpc & \\
$r_{p,1}$ (simulation orbit) & (0.35)/3.5 kpc & \\
r$_{\rm disk}$	(sim)/physical	  	&(1/12)/0.835 kpc	& (1/33.941)/0.295 kpc\\
z$_{\rm disk}$ (sim)/physical	  	& (0.0125)/ 0.125 kpc 	& (0.00442)/0.044 kpc  \\
a$_{\rm halo}$ (sim)/physical	  		 & (0.25)/2.505 kpc &(0.0884)/0.886 kpc	\\
b$_{\rm halo}$ (sim)/physical	  		 & (0.975)/ 9.77 kpc & (0.3448)/ 3.45 kpc	\\
c$_{\rm halo}$	  		 &3.9 & 3.9	\\
($i,\omega$)-disks		&(58 $\pm 5\deg$,115 $\pm 20 \deg$) 	&(32 $\pm  10\deg$,92 $\pm 15\deg$) 	\\
$\Delta v_{3D}$ 		& $\sim 85$ \kms  &\\
$\Delta pos_{3D}$	& $\sim 28$ kpc&  \\
	& &\\
{\bf \large t$_{match}$} (present day)	& &\\
\hline
M$_{\rm disk, baryons}$($< 7\times r_s$) 	 & 3.76 $\times 10^9$ M$_{\odot} $&0.33 $\times 10^9$ M$_{\odot} $ 	\\
$\Delta v_{\rm 3D}$ 		& 43 \kms~ &\\
$\Delta pos_{\rm 3D}$ 	& 9.3 kpc  &  \\
$\Delta pos_{\rm projected}$ 	& 6.5 kpc  &  \\
Viewing angles &	 (87$\pm 3 \deg$, 309$\pm 3 \deg$, 55$\pm 3 \deg$)& \\
Galaxy inclination sky view&	70.0 $ \pm 15\deg$& 88 $ \pm 10\deg$ \\
\hline
\end{tabular}
\end{table*}

\subsection{The dynamical match to NGC 4490/85 }\label{sec:match}
Our best test-particle simulation match to the data of NGC 4490/4485 has a galaxy mass ratio of $\mu = 8:1$, and a first pericentric separation of $r_{peri} = 4.2 \times r_{s,prim}$ (3.5 kpc). 

Subsequent to obtaining the best test-particle simulation match, we run three $N$-body follow-up simulations with self-gravitating disks and pericentric separations close to the value for the best test-particle match ($r_{peri} = 3.4, 4.2, 5.0 \times r_{s,prim}$).
When we load these into the \identikit~ visualization interface, we find that for the three $N$-body follow-ups, the best agreement with the data was still for the $r_{peri} = 4.2 \times r_{s,prim}$ case, with minor variation from the test-particle match by a few degrees in the viewing angles of the system. 
We therefore present and analyse only the $N$-body follow-up with $r_{peri} = 4.2 \times r_{s,prim}$ throughout this section. In Table \ref{tab:model} we present the specific scalings and galaxy properties obtained in our match. 
The $t_{start}$-column describes the initial conditions with the scaling values from the best-match applied. The $t_{match}$ column describes the properties at the time of match (i.e. present day, $t = 0$ Gyr). 

For our best $N$-body match, the two galaxies are separated by 28 kpc at the beginning of the simulation (t$_{start}$ = -1.528 Gyr) and initially follow a Keplerian ($e = 1$) orbit (see Figure \ref{fig:orbit}). The idealized Keplerian orbit has an initial pericentric separation of $r_{p,1} = 2.5$ kpc, however the first pericentric passage using the extended galaxies from the $N$-body and test-particle simulations deviates slightly from the idealized Keplerian orbit, resulting in a pericentric approach of $r_{p,1} = 3.5$ kpc (see Figure \ref{fig:orbit}). 

The first pericentric passage occurs $\sim 0.2$ Gyr after the start of the simulation and the orbit decays due to dynamical friction between the galaxies. The time of match ($t = 0$ Gyr) 
occurs between the second and third pericentric passage, close to apocenter, and is 1.29 Gyr after the first pericentric passage between the two galaxies. 
The viewing angle that affords the best match is aligned with the orbital plane. Hence from our perspective, the tidal tails are aligned along our line-of-sight (see Figure \ref{fig:match}).
In the following we discuss the  orbital solution of the match (Section \ref{sec:orbit}), the morphology of the match (Section \ref{sec:morph}), the kinematics of the match (Section \ref{sec:kin}) and the initial properties of the primary and secondary galaxy (Section \ref{sec:prop}) in detail.

\subsubsection{The orbit}\label{sec:orbit}
\begin{figure*}
\centerline{\includegraphics[width=\textwidth]{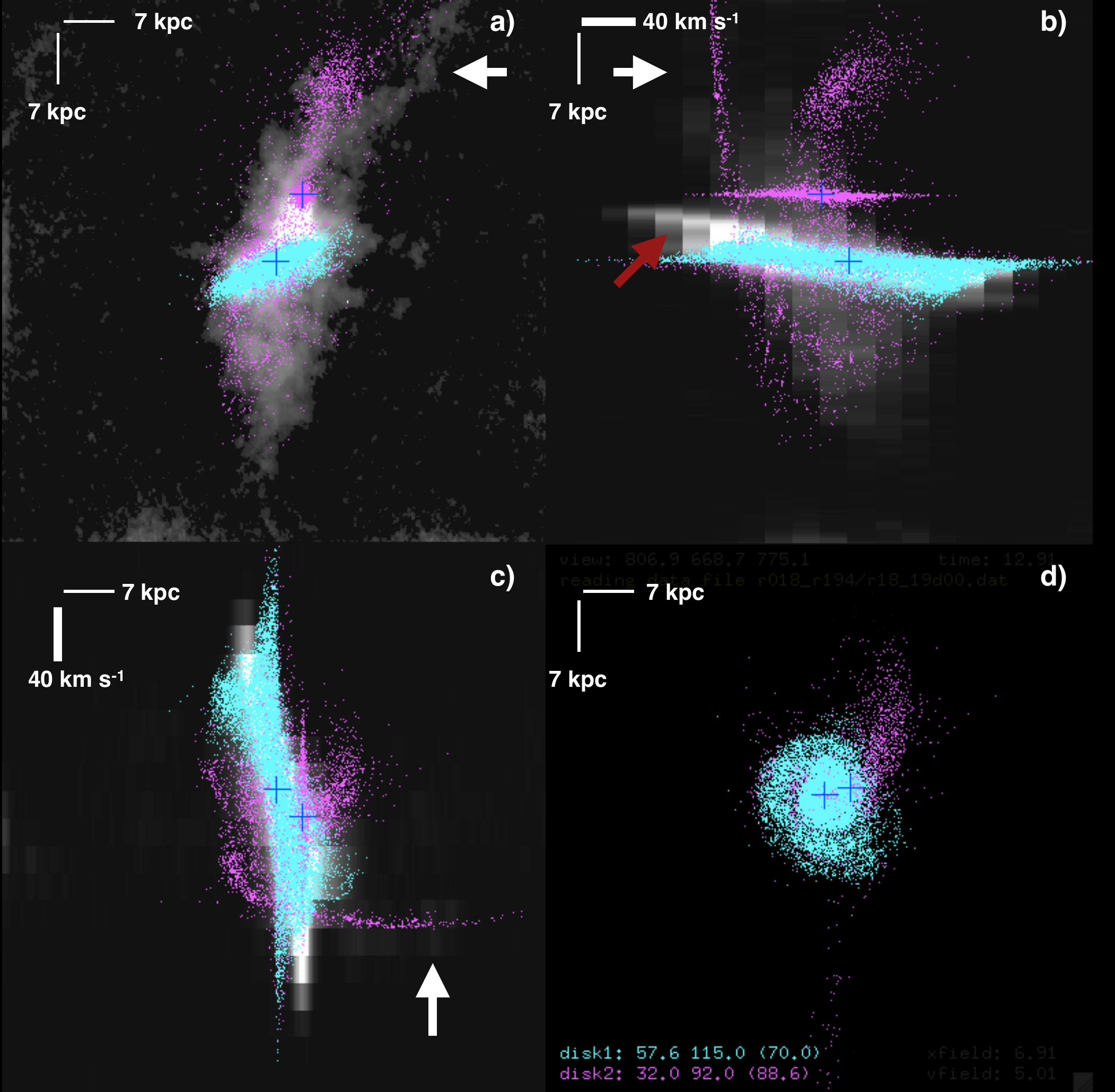}}
\caption{Visualization of the self-consistent $N$-body simulation for NGC 4490 (cyan)  and NGC 4485 (magenta), matched to the observed system based on \identikit~ test-particle simulations. The match quantitatively reproduces the kinematics and morphology for galaxy values that mimic the system. (a) sky view of the system (RA-Dec), (b) line-of-sight velocity vs position diagram (vel-Dec), (c) position vs line-of-sight velocity diagram (RA-vel), and (d)``top-down" (RA vs line-of-sight distance) view of the simulation. Assuming a distance of 7.14 Mpc to the system (\citealt{theureau07}) the sky view covers 69.2 kpc $\times$ 69.2 kpc.
The HI data from \citet{clemens98} is shown in grayscale, with the lighter pixels corresponding to higher peak values along a vector through the data cube. The velocity range is $-123$ \kms~ to 83 \kms~ with a velocity width per channel of 20.7 \kms, and the velocity increases from left to right in panel b and from bottom to top in panel c.
The cyan and magenta points show collisionless baryonic particles from the self-gravitating primary and secondary representing the galaxies NGC 4490 and NGC 4485, respectively. The blue crosses represent the nuclei of each $N$-body realization. The match shown here occurs between the second and third pericentric passage, which is $\sim$1.29 Gyr after the first pericentric passage and $\sim$230 Myr after the second passage. The white arrows point to the end of the secondary's tidal tail produced in the first pericentric passage and the red arrow points to the HI emission in the bridge.
}
\label{fig:match}
\end{figure*}

The secondary galaxy has a prograde spin with respect to the orbital angular momentum vector enabling substantial mass loss such that the envelope is mainly produced from material from the secondary galaxy (see magenta particles in  Figure \ref{fig:match}). 
In Figure \ref{fig:orbit} we show the evolution of the primary and secondary galaxies' nuclear separation (left), velocity separation (middle) and the secondary's orbital evolution around the primary galaxy (right) for the $N$-body simulation of the match shown in Figure \ref{fig:match}. The color bar illustrates  the time evolution with respect to present day, which is the time of the best match in the simulation ($t = 0$ Gyr). 
The physical  scaling is based on the best match (see Table \ref{tab:model}) and assuming a distance of 7.14 Mpc (\citealt{theureau07}) to the system.  At the time of match (red solid line), two close encounters have occurred ($r_{p,1} = 3.5$ kpc, $r_{p,2} = 1.7$ kpc) and the primary galaxy's tidal field strips material from the secondary at each close encounter. 

The fact that the match (red solid line) occurs close to apocenter, where the two galaxies are farthest apart, is not surprising as this is the point in their orbit at which they spend the most time (i.e. move at the lowest velocity, see Figure \ref{fig:orbit} middle panel). In Figure \ref{fig:orbit}, left, we show that the galaxies are separated by $d_{3D} =$ 9.3 kpc ($d_{\rm proj} =$ 6.5 kpc) at the time of match and that the relative velocity between the two nuclei is 43 km s$^{-1}$ (Figure \ref{fig:orbit}, middle) in good agreement with observations (see Table \ref{tab:prop}). 

The time between the first and second pass is $\sim$1.1 Gyr, and following the simulation after the time of match reveals the prediction that the dwarfs coalesce $\sim 1.7$ Gyr after the first pericentric passage (370 Myr after the time of match: Figure \ref{fig:orbit}, left, gray vertical line). Thus,  due to the large mass ratio between the two galaxies and their orbital configuration capturing the two galaxies separately as an interacting pair is not a short-lived stage (see also \citealt{besla16}). Similar merger timescales are also found for high mass ratio encounters for massive galaxies (see e.g.  \citealt{cox08}, \citealt{patton13}, \citealt{jiang14}). 
 
Given the decay time of $\sim$ 1.7 Gyr, which is quite rapid compared to a Hubble time, our best match orbit indicates that the two galaxies likely formed very far apart and have had a long infall time as we still observe the two galaxies separately today. This provides dynamical insight to the survivability of these types of of low mass galaxy pairs and groups (e.g. \citealt{stierwalt17}). Our study suggests that close pairs viewed today are likely to have begun their encounter on high eccentricity orbits to prevent rapid merging and are unlikely to have survived as bound systems for a Hubble time. 

The third panel of Figure \ref{fig:orbit} shows the evolution of the secondary's orbit around the primary in the plane of the sky. The orbit is centered on the primary galaxy, the initial position of the secondary is marked by the magenta star and the red star indicates the position of the secondary at the time of match. From our viewing perspective the secondary galaxy is initially on a high inclination orbit with respect to the disk plane of the more massive galaxy, spanning $\sim$ 30 kpc in the Declination direction and only $\sim$ 15 kpc in the Right Ascension direction.

We stress that when exploring the \identikit~ library, we found that the extended morphology of the envelope could generically be reproduced by this broad type of orbital configuration. However reproducing the details of the match and its kinematics required narrowing down the free parameters of the disks (see Section \ref{sec:kin}).

\subsubsection{The morphology of the preferred match}\label{sec:morph}
Figure \ref{fig:match}, panel {\bf a)} shows the morphology of the observed (grey scale) and simulated data (colored points) in RA-Dec. 
The magenta particles (secondary galaxy, NGC 4485 analog) in Figure \ref{fig:match},  panel {\bf a)} populate both the north and the south of the data morphologically. The magenta particles are partially made up of a long tidal tail stripped from the secondary on its first pericentric passage which wraps around the primary galaxy as a roughly symmetric 50 kpc envelope when viewed from our perspective. Hence, the viewing angle places the tail stripped in the first pass mostly along the line-of-sight, populating both the northern and southern envelope, which explains why the tidal tails do not look like long thin features (e.g. as the case for the Antennae galaxies: \citealt{toomre72}).
The fact that the match occurs between the second and third pericentric passage, allows for the tidal tail from the secondary galaxy's first pericentric pass to grow and populate the full extent of the $\sim 50$ kpc HI envelope. Additional material stripped at the second pericentric passage (see material north of the magenta main body, Figure \ref{fig:match}  panel {\bf a}) also contributes to the HI envelope. 

\subsubsection{The kinematics of the preferred match}\label{sec:kin}
Figure \ref{fig:match}, panel {\bf b)} shows line-of-sight velocity vs. position (vel-Dec) with increasing velocity from left to right. Panel {\bf c)} shows the position vs velocity (RA-vel), where the velocity is increasing from bottom to top. The data and magenta particles in the tail feature in the lower right part of panel {\bf c)} is marked by a white arrow in each of panels {\bf a}, {\bf b}, and {\bf c}, although the contrast in Figure \ref{fig:match} does not highlight the observational data clearly in this region. This feature is the end of the tail produced in the first pericentric passage, which wraps around the system when viewed from our perspective (see Section \ref{sec:morph}). This structure populates both the north and the south of the envelope morphologically and kinematically. To reproduce this ``end-of-the-tail" feature, a specific configuration was required: viewing angles $\theta_x = 87\deg$ , $\theta_y = 309 \deg$, $\theta_z = 55\deg$ and secondary disk orientation: $i_{sec} = 32\deg$, $\omega_{sec}=92\deg$. We explored the acceptable range in the viewing angles by varying them in the $N$-body follow-up until the simulated particles no longer provided a good match to the system. Via this approach we found that the approximate uncertainty on each viewing angle is $\pm 3\deg$. We carried out the same test for the disk orientation using the test-particle simulation (we cannot vary the disk inclinations in the $N$-body follow-up, as they are modeled self-consistently) and found that the approximate uncertainty was $i_{sec} = 32 \pm 10\deg$ and  $\omega_{sec} = 92 \pm 15\deg$. 

To match the morphological and kinematic gradient along the primary's body while matching the extended tidal debris required a specific disk orientation ($i_{prim}=58\deg$, $\omega_{prim}=115\deg$) for the primary. We explored the acceptable range in the primary disk orientation by varying $i_{prim}$, $\omega_{prim}$ in the test-particle simulation until it no longer provided a good match to the system. Via this approach we found that the approximate uncertainty on the primary disk orientation is $i_{prim} = 58 \pm 5\deg$ and  $\omega_{prim} = 115 \pm 20\deg$.

Observational data marked by the red arrow in Figure \ref{fig:match}, panel {\bf b} are unpopulated by magenta or cyan particles. This observational feature corresponds to  the higher column density bridge material at $v\sim -123$ to $-82$ \kms~ (see Section \ref{sec:hi}). We see the secondary move through the disk of the primary during the most recent passage, but this feature likely is not reproduced due to the lack of hydrodynamics in our simulations (see Section \ref{sec:test} and Section \ref{sec:hydro}). As pointed out by \citet{clemens00}, gravitational forces do not distinguish between gas and stars in a galaxy encounter, and that NGC 4485 could experience ram pressure stripping as it moves through the extended HI distribution. In our model, NGC 4485 passes through the gaseous disk of NGC 4490, and therefore will experience significant ram pressure stripping during this time. More gas would be thus be stripped than by tides alone. This is also the theory to explain why the Magellanic Bridge has a higher gas density than the Magellanic Stream (\citealt{besla12}). \citet{clemens00} additionally pointed out that this picture is consistent with the fact that the HI distribution of NGC 4485 is offset from its optical counterpart, and that there is evidence of a bow-shock ahead of the stripped gas. In future work, we plan to test the hypothesis that the gaseous bridge can be reproduced by including hydrodynamical effects in our simulations (see also Section \ref{sec:hydro}).

We use the optical centers as additional constraints on the model. The secondary galaxy (magenta particles) is therefore located at  $v\sim -20$ \kms~ and the primary galaxy is centered at $v\sim 0$ \kms~ (see Section \ref{sec:hi}). We reproduce both the kinematic position of the primary and secondary in panel {\bf b} and {\bf c}.

\begin{figure*}
\centerline{\includegraphics[width=\textwidth]{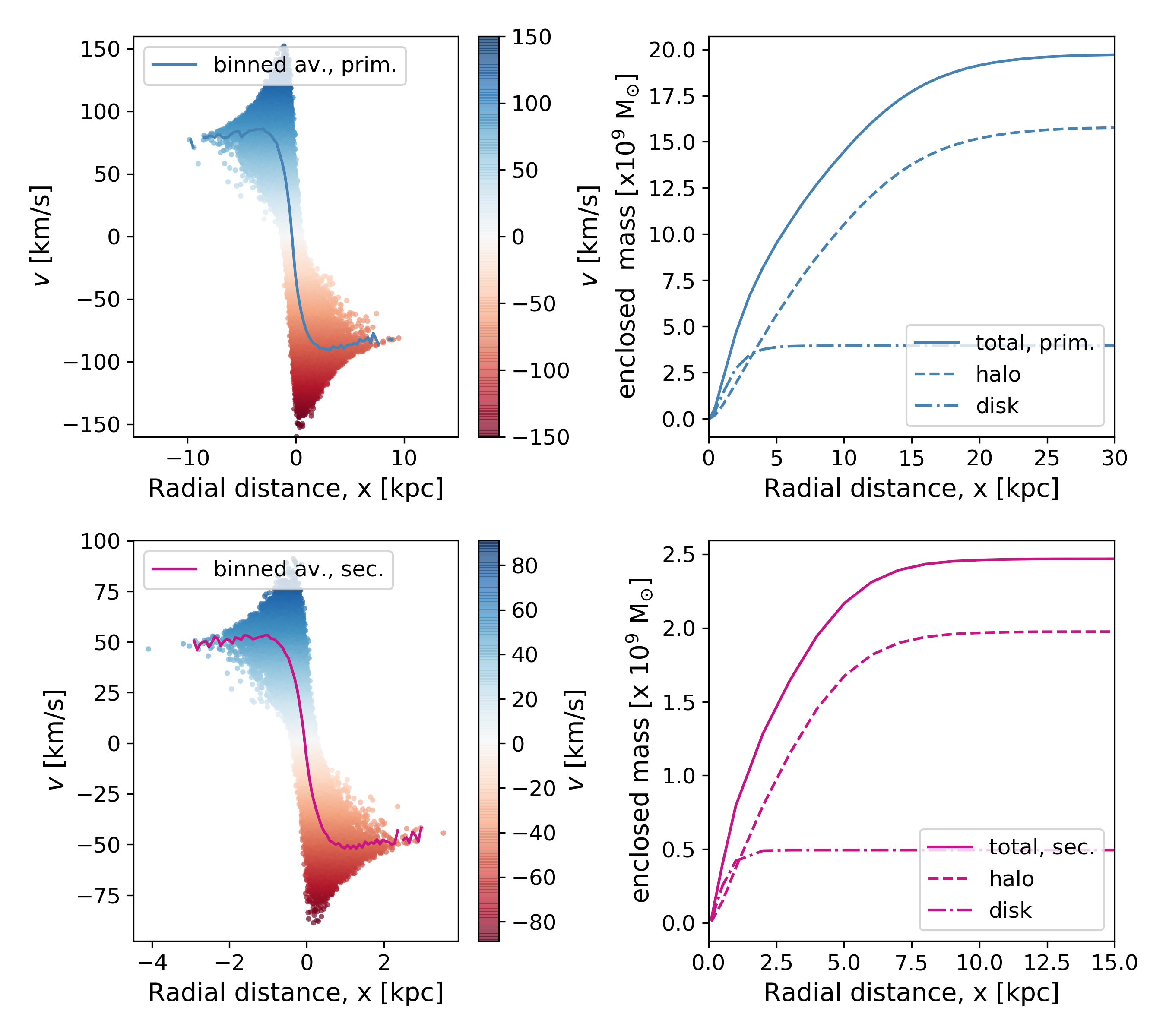}}
\caption{{\bf Left:}  Velocities of our simulated primary (top) and secondary (bottom) galaxies as a function of radial distance from their centers at the beginning of our simulation ($t_{start}$). The size and velocity scale is derived from our mass models, using the best match parameters (see Table \ref{tab:model}). The color coding shows the velocities in the edge on projection of the two galaxies.The blue (primary) and magenta (secondary) lines show the averaged $v$, binned in 100 bins. Our primary galaxy model has a peak velocity of $v_{\rm peak} \sim 80 ~\kms$ which then flattens after 4 kpc. This is roughly consistent with the observations of NGC 4490: at present day NCC 4490's peak rotational velocity is $v_{\rm rot,peak} \sim 80 ~\kms$ (\citealt{elm98}) whereafter the rotation curve drops and does not follow a flat curve. The secondary's rotation curve is similar to that of the SMC (\citealt{stani04}) peaking at $v_{\rm rot,peak} \sim 50 ~\kms$ and does not have an observationally derived curve. 
{\bf Right}: Spherically averaged enclosed baryonic and dark mass profiles of the primary (top) and secondary (bottom) galaxy at $t_{start}$ scaled based on our best match parameters (see Table \ref{tab:model}).  }
\label{fig:rot}
\end{figure*}

\subsubsection{Initial properties of the primary and secondary}\label{sec:prop}
Throughout the paper we have used mass models mimicking those of massive galaxies (see Section \ref{sec:massmodel}, \citealt{barnes09}). In Figure \ref{fig:rot}, left we show the disk rotation curve and the mass profiles of our primary and secondary galaxy models at the beginning of our simulation $t_{start}$ scaled based on the best match parameters (see Table \ref{tab:model}). For the primary galaxy, our rotation curve peaks at $v_{\rm rot,peak} \sim 80$ \kms, which is the same value as obtained for the observational HI rotation curve (\citealt{elm98}). We used the kinematic extent of the data as part of the \identikit~ matching procedure  (see Figure \ref{fig:match} panel b, c), so it is encouraging that our initial conditions based on our preferred match are consistent with the observed present day rotation curve for NGC 4490.
Our primary galaxy's rotation curve peaks at slightly smaller radius than what \citet{elm98} found (6 kpc), which could be addressed with a different mass model. There is no observational rotation curve available for the secondary galaxy, and we expect hydrodynamical effects to have distorted the present day HI within NGC 4485 as it has recently passed through NGC 4490 (see discussion in \ref{sec:hydro}). Our secondary's simulated rotation curve peaks at $v_{\rm rot,peak} \sim 50$ \kms~ in the beginning of the simulation, which is similar to that of the SMC ($v_{\rm rot,peak} \sim 60$ \kms~ at 3 kpc: \citealt{stani04}) if it is modeled.

Based on our match the implied total initial baryon mass of the primary and secondary in the simulation are $m_{\rm disk,prim} = 4 \times 10^9 ~\msun$ and $m_{\rm disk,sec} = 0.5 \times 10^9 ~\msun$, respectively (see Table \ref{tab:model} and Figure \ref{fig:rot}, right). This initial baryonic mass is a factor of $\sim 2.6$ lower compared to the total baryon mass in the data (see stellar masses, HI masses and envelope HI mass in Table \ref{tab:prop}).
Obtaining a better match to the NGC 4490/4485 system would require carefully altering the mass models by making the halos less concentrated and then adding mass back in baryons, which is beyond the scope of this paper. We stress that where we can compare to data (i.e. the NGC 4490 rotation curve), our simulated galaxy mass profiles are consistent with observations (see rotation curves Figure \ref{fig:rot}, and morphological and kinematic match in Figure \ref{fig:match}).  

Compared to what we would expect from abundance matching (e.g. \citealt{moster13}) the halo masses for our two galaxies are quite low: $m_{\rm halo,prim} = 1.6 \times 10^{10} ~\msun$ and $m_{\rm halo,sec} = 2 \times 10^9 ~\msun$, respectively (Table \ref{tab:model} and total enclosed mass in Figure \ref{fig:rot}, right). The observed stellar mass ratio is 8:1, but in our simulation both the stellar mass ratio and halo mass ratios are 8:1. If the two galaxies were isolated, from abundance matching (using Eq. 2 in \citealt{moster13}) we obtain a halo mass of  $m_{\rm halo} = 2.6 \times 10^{11}$ ~\msun~ for NGC 4490 and $m_{\rm halo} = 9.1 \times 10^{10}$ ~\msun~ for NGC 4485 (see Table \ref{tab:prop}). This corresponds to a halo mass ratio of $\sim 2.85:1$ although there is a large scatter in the \citet{moster13} abundance matching relation at small stellar masses. 
The dark matter halo mass at large radii is therefore not well constrained by our simulation match, 
and it is possible that more dark matter mass is present at larger radii than in our galaxy mass models presented here (Figure \ref{fig:rot}, right).

As a sanity check of the robustness of our retrieved encounter geometry, we explored mass models with galaxies more consistent with expectations from $\Lambda$CDM with less baryons as compared to dark matter ($\sim$3\% baryons as opposed to 20\%), and larger halo concentrations ($c_{halo}  = 12$ instead of $c_{halo}  = 3.9$). We found that using the same viewing angles, disk angles, time of match and initial pericentric separations the match did not significantly change although this resulting match favored more massive halos (by a factor of 3.5). Hence the encounter geometry seems robust to the specific choice of mass model. 
In Section \ref{sec:rate} we discuss the consequences of our lower dark halo mass models when we explore the fate and return timescales of the envelope. 

See the {\it Appendix} for a discussion of alternative matches which reproduce the character of the system, but do not provide as satisfactory matches.

\subsection{The Formation of the Extended HI Envelope}\label{sec:baryon}
\begin{figure*}
\centerline{\includegraphics[width=\textwidth]{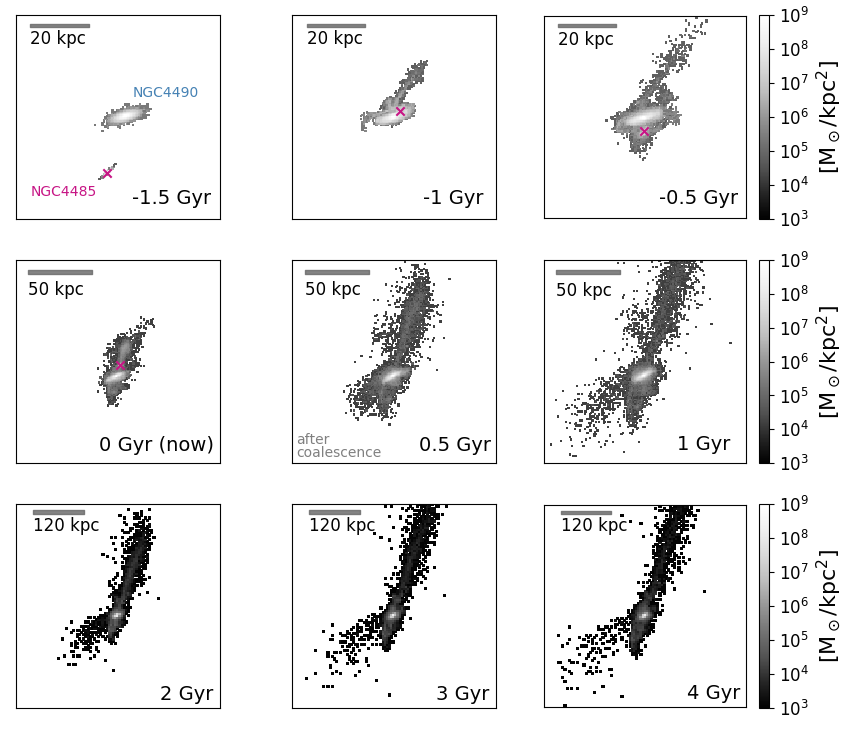}}
\caption{Snapshots of the evolution of the collisionless baryonic material from the $N$-body simulation of the bast match to NGC 4490/85 as a function of time. The time is indicated relative to the time of match (present day, $t = 0$ Gyr). The magenta x's indicate the center of the secondary galaxy (NGC 4485) prior to its disruption. The panels are centered on the primary galaxy. The pair coalesces between $t = 0$ and $t = 0.5$ Gyr, after which the debris continues to grow in size and persists for several Gyr. The color bar denotes the density of the material, which is small in the large envelope compared to the densities in the main bodies at the time of match. The densities are converted to physical units based on the particle masses and the bin sizes in each row, and we saturate the density at $10^9$ \msun~ kpc$^{-2}$to better illustrate the faint features. The scale bars and densities are plotted assuming a distance of 7.14 Mpc (\citealt{theureau07}). Note the difference in the spatial scaling between the three rows.}
\label{fig:privon13}
\end{figure*}

The match presented in the previous subsection demonstrates that the extended $\sim 50$ kpc HI envelope can be reproduced through a dynamically-driven dwarf-dwarf tidal encounter formation scenario. While outflows might serve to add more mass to the envelope, our results show that a mutual interaction between two dwarf galaxies can move baryons to very large distances through tides alone, without the presence of outflows, tidal stripping through the Lagrange points due to a host galaxy's tides (e.g. a Milky Way) or ram-pressure effects.
In this section we investigate the formation of the extended envelope and the mass and tail evolution of the tidal debris (see Sections \ref{sec:mass} and \ref{sec:stream}). Throughout the rest of the paper we alternate between showing the system in the plane of the sky and in a ``top-down" view (RA vs line-of-sight distance as in Figure \ref{fig:match}, panel {\bf d}). This enables us to highlight various features of the extended tidal debris. 

Figure \ref{fig:privon13}  illustrates the evolution of the baryonic density distribution of the NGC 4490/4485 system throughout their encounter, projected along our line-of-sight. Initially the two galaxies are separated by $d_{3D}$ = 28 kpc. The first pericentric passage occurs at $t =-1.29$ Gyr (between the $-1.5$ and $-1.0$ Gyr panels). In the first three panels we see the formation of the envelope and the subsequent panels show its predicted evolution. From our viewing perspective the debris stripped from the secondary in the first pericentric passage wraps around the primary's disk. 

The middle, left panel of Figure \ref{fig:privon13} shows the system at the time of match, where a large 50 kpc (projected) envelope surrounds the two dwarf galaxies which are separated by $d_{3D} =$ 9.3 kpc corresponding to a projected separation of $d_{\rm proj} =$ 6.5 kpc. The surface density of material in the large envelope is a factor of $10^4$ lower than the densities in the main bodies, and the tidal debris in the north and south of the envelope is of similar surface density, which is consistent with the \citet{clemens98} observations. 

Figure \ref{fig:privon13} additionally shows that the debris continues to grow in size throughout the encounter and the bulk of the material does not immediately return to the system. In particular, the debris field will persist and should be observable long after the system coalesces (at $\sim$370 Myr). In the last snapshot we see that the final system will look like a companionless galaxy surrounded by a system of gaseous streams, although the exact properties of the merger remnant will depend on dissipational effects which are not included in this analysis.

\begin{figure}
\centerline{\includegraphics[width=\columnwidth]{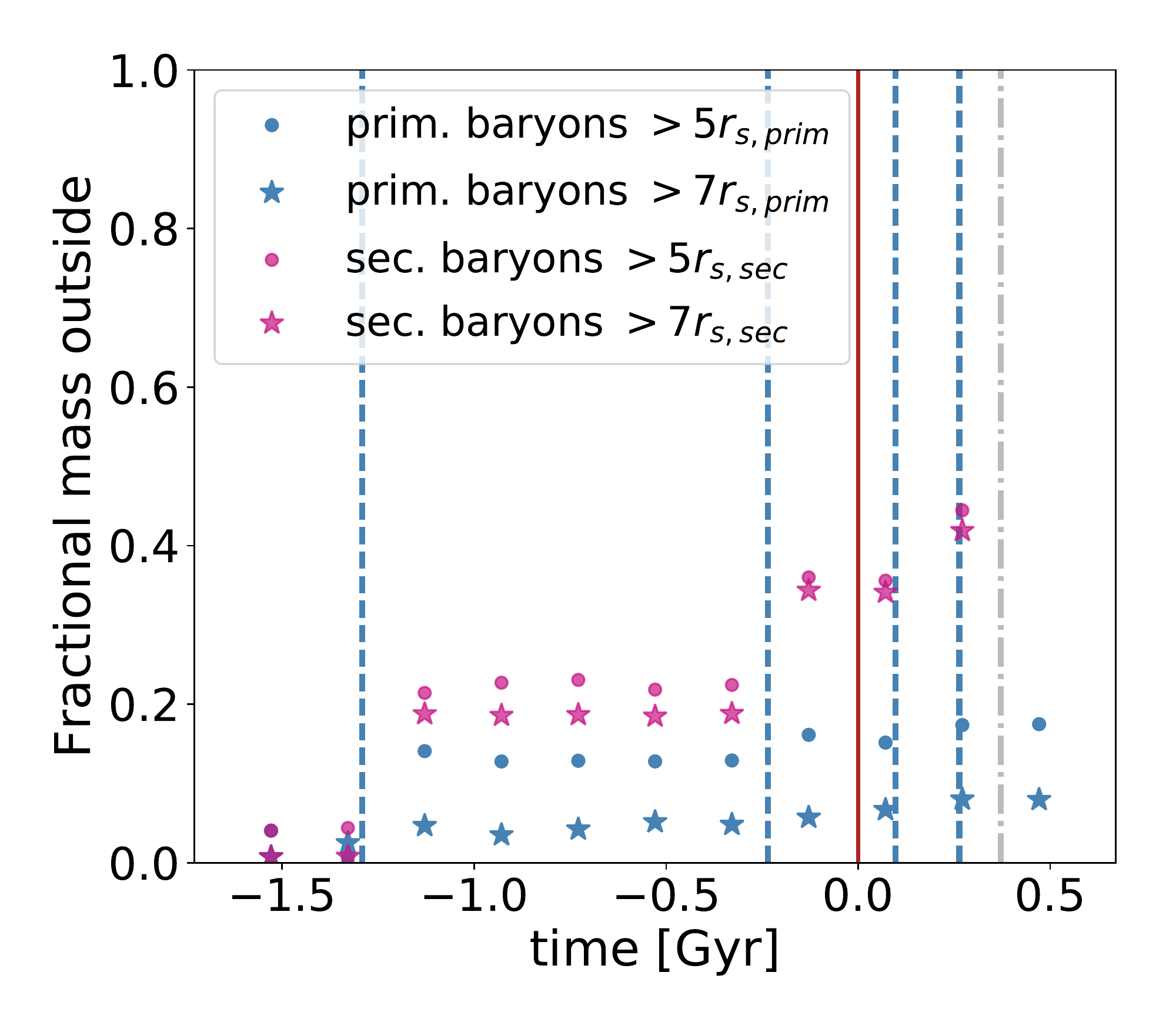}}
\caption{The fraction of baryonic mass residing outside 5 (circle markers) and 7 (star markers) disk scale radii divided by the total initial baryonic mass of each galaxy is plotted as a function of time, in 200 Myr increments.  Blue points indicate results for the primary and magenta points for the secondary. We do not account for mass transfer between the two galaxies in this figure. The vertical blue dashed lines correspond to pericentric passages, the red solid line indicates the time of match (present day) and the gray dash dotted vertical line demonstrates when the two galaxies coalesce, after which we do not track the mass outside the secondary. At the beginning of the simulation all baryonic material resides within 7$r_s$ of each galaxy respectively (see star markers). After each pericentric passage both galaxies get more extended and the secondary loses a substantial fraction of its mass. Given the eccentricity of the orbit, tidal stripping does not proceed smoothly over the course of the encounter.  After coalescence the fractional masses outside the main bodies remain constant until re-accretion of material (see Figure \ref{fig:rate}). }
\label{fig:moutside}
\end{figure}

\subsubsection{Evolution of the Baryonic Disk Mass Distribution}\label{sec:mass}
To quantify how much baryonic material is pulled out in tidal structures during this encounter, in Figure \ref{fig:moutside} we show the fraction of the baryonic particles removed from the primary and secondary by their mutual tidal forces as a function of time. 
The two galaxies are investigated separately, and we do not account for transfer of mass between the two galaxies in this plot. Instead, at each time step we evaluate the amount of mass beyond the listed radius (5 or 7 $\times$ the scale radius) of each galaxy separately. 

Initially (at $t = -1.53$ Gyr), all baryons reside within 7$r_s$, where $r_s$ is the disk scale length of each galaxy, listed in Table \ref{tab:model}.  After each pericentric passage both galaxies become more extended and/or have baryons pulled out in tidal features. Fractionally, the secondary loses much more material than the primary at each pericentric passage in the simulation (see Figure \ref{fig:moutside}, magenta vs cyan points). This is due to the high mass ratio between the two galaxies and the fact that the secondary is spinning prograde with respect to its orbit around the primary (e.g. \citealt{toomre72}, \citealt{dongia10}, \citealt{Sengupta15}). While some of the secondary's particles do end up within 7 disk scale radii of the primary ($<5\%$ of the secondary's particles after both the first and second pericentric pass), the point here is to track which galaxy loses the most mass fractionally in the encounter.

The mass loss occurs at pericenter, with little evolution in the mass profile between pericenters. At the time of match (Figure \ref{fig:moutside}, red vertical line) $\sim$ 34\% of the baryons from the secondary galaxy reside outside $7r_{s,sec}$,  
and only $\sim$ 6\% of the primary's baryons are beyond 7$r_{s,prim}$. 
Interestingly, the difference between the amount of material within 5$r_s$ and 7$r_s$ at a given time step is larger for the primary galaxy than for the secondary after the first pericentric pass (the blue circles vs. stars are more separated than the magenta circles vs. stars). This indicates that the primary is being morphologically affected by the secondary, despite the high mass ratio of $\mu = 8:1$ (see Section \ref{sec:prim} for a discussion of this).

Observationally, the amount of HI gas outside the stellar disks of NGC 4490/4485 is M$_{\rm HI}$(outside)$ = 1.07 \times 10^9 ~\msun$ (as defined in \citealt{pearson16}, table 2, 3 and shown as red ellipses in their figure 1). This estimate was done defining the size of the stellar disks of NGC 4490/4485 as 4 times the $K_s$-band scale length from the 2MASS catalog. This corresponds to $r_{\rm ext,prim} = 7.4$ kpc and $r_{\rm ext,sec} = 2.6$ kpc for NGC 4490/4485 respectively\footnote{Note that 4 times the $K_s$-band scale length from the 2MASS catalog yields extents which are similar to the $R_{25}$ extents of both galaxies: $R_{25} = 6.5$ kpc (\citealt{elm98}) and $R_{25} = 1.1$ kpc (The NASA/IPAC Extragalactic Database, NED) for NGC 4490 and NGC 4485, respectively assuming a distance to the pair of 7.14 Mpc.}, which is slightly larger than but comparable to 7 disk scale lengths used for our analysis in this paper (7$r_{s,prim} = 5.8$ kpc and $7r_{s,sec} = 2.1$ kpc). 
In our simulation at the time of match the amount of baryons beyond 7 disk scale radii of the primary and secondary galaxies correspond to  $6.9 ~\times~ 10^8 ~\msun$ and  $ 2.3  ~\times~ 10^8 ~\msun$, respectively (see Table \ref{tab:model}).  Hence in our match, a total baryon mass of $ \sim 0.9  ~\times~ 10^8 ~\msun$  resides outside the galaxies at the time of match, which is similar to the observed value for M$_{\rm HI}$ outside the 2MASS extents of the galaxies. 

While it is encouraging that this number is the right order of magnitude, we do not distinguish between stars and gas in our baryon mass budget. In this section we are comparing to HI observations, but stars should similarly be tidally removed in the interaction. As mentioned  in Section \ref{sec:prop}, varying the mass models (e.g. the disk extent, halo concentration, halo mass) could change the amount of mass lost from the baryonic disks during the encounter. In this work we have matched the simulations to HI data, and typically gas disks are more extended than stellar disks (\citealt{swaters02}) which could ensure that the tidal debris would be primarily made of gas, rather than stars. This is the explanation commonly invoked to explain the lack of stars in the Magellanic Stream (\citealt{besla12}).

\subsubsection{Visualizing the Tidal Tail Debris}\label{sec:stream}
\begin{figure*}
\centerline{\includegraphics[width=\textwidth]{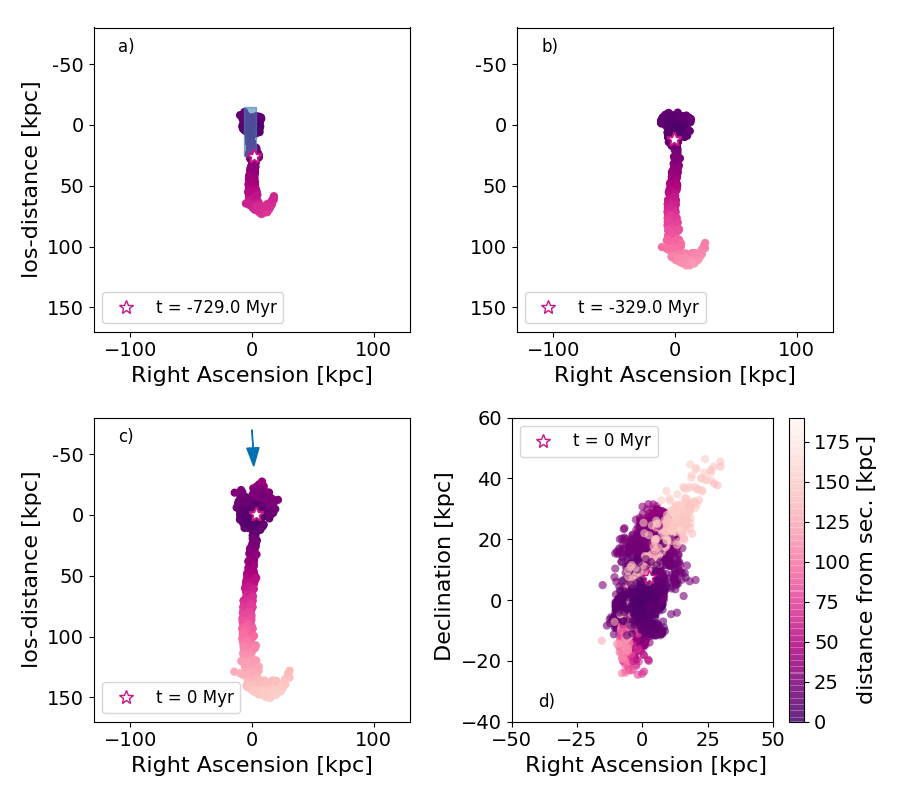}}
\caption{{\bf Panel a through c:} ``top-down" view (RA vs line-of-sight distance) of the evolution of the secondary dwarf's (NGC 4485) baryonic material beyond 7 disk scale radii at three different time steps. Panel {\bf a)} is roughly the time of first apocenter,  Panel {\bf b)} is close to the time of second pericentric passage, Panel {\bf c)} and {\bf d)} is at the time of match. Each plot is centered on the position of the primary dwarf (NGC 4490) (x,y = 0,0). The star indicates the position of the secondary at each snapshot. The blue box in panel {\bf a)} indicates the physical extent of the secondary's orbit. The blue arrow in panel {\bf c)} indicates the skyview angle for an observer at the location of the Earth. The color demonstrates the 3D distance to each particle from the center of the secondary. {\bf Panel d}: A rotation of panel {\bf c)} to skyview showing all the secondary's particles beyond 7 disk scale radii at the time of match ($t = 0$ Myr). The tail lost at the first pericentric passage continues to grow beyond the size scale of the orbit. Due to our viewing angle of the system, we see the tail as a 50 kpc symmetric envelope surrounding the dwarf galaxy pair.}
\label{fig:tailevol}
\end{figure*}

To explore in detail how the tidally removed material evolves relative to the galaxies, in Figure \ref{fig:tailevol} we investigate the ``top-down" view (RA vs line-of-sight distance) of the simulation (see blue arrow in panel {\bf c} for the observers line-of-sight view, which would recover the perspective plotted in Figure \ref{fig:privon13}). In particular, we show the morphological evolution of the particles beyond 7 disk scale radii of the secondary dwarf (see magenta stars in Figure  \ref{fig:moutside}) at three different times (Figure \ref{fig:tailevol}, panel {\bf a} through {\bf c}). To ensure that the tidal features have had time to grow, the first time step is near the first apocenter (see Figure \ref{fig:orbit}, left). The color scale indicates the 3D distance of each particle from the center of the secondary galaxy at each time step. In this projection, the orbit of the secondary around the primary is confined within the blue box in the first panel and is shown here for scale. The magenta star indicates the position of the secondary in each panel, and the first three panels are centered on the position of the primary dwarf.

Figure \ref{fig:tailevol} shows that the tidal tail produced in the first passage grows $> 100$ kpc beyond the physical scale of the orbit. As expected, the tail does not stay in the plane of the orbit due to the offset between the orbital plane and the inclination of the secondary dwarf. Panel {\bf c)} shows the time of match ``top-down" view and the blue arrow indicates our viewing perespective of the system. At the time of match the tail initially raised at the first pericentric passage has grown to be $\sim$175 kpc in size. Interestingly, the ongoing Arp 299 galaxy merger (\citealt{hibbard99}) has one of the longest HI tails observed ($\sim$180 kpc in projection) and is a factor of 10 more massive than the NGC 4490/4485 system. The true 3D length of Arp 299 might be even longer than 180 kpc due to projection effects and the sensitivity of the observations. The fact that the dwarf encounter between NGC 4490/4485 produces a tail of similar length emphasizes that low mass dwarf-dwarf encounters can have dramatic effects, similar to the tidal distortions induced by interactions between massive galaxies. When we rotate the system into the plane of the sky (Figure \ref{fig:tailevol}, panel {\bf d}), the long tail wraps around the primary galaxy as a roughly symmetric $\sim$50 kpc envelope when viewed from the Earth (as described in Section \ref{sec:morph}).

\subsection{Consequences for the primary galaxy}\label{sec:prim}
\begin{figure*}
\centerline{\includegraphics[width=\textwidth]{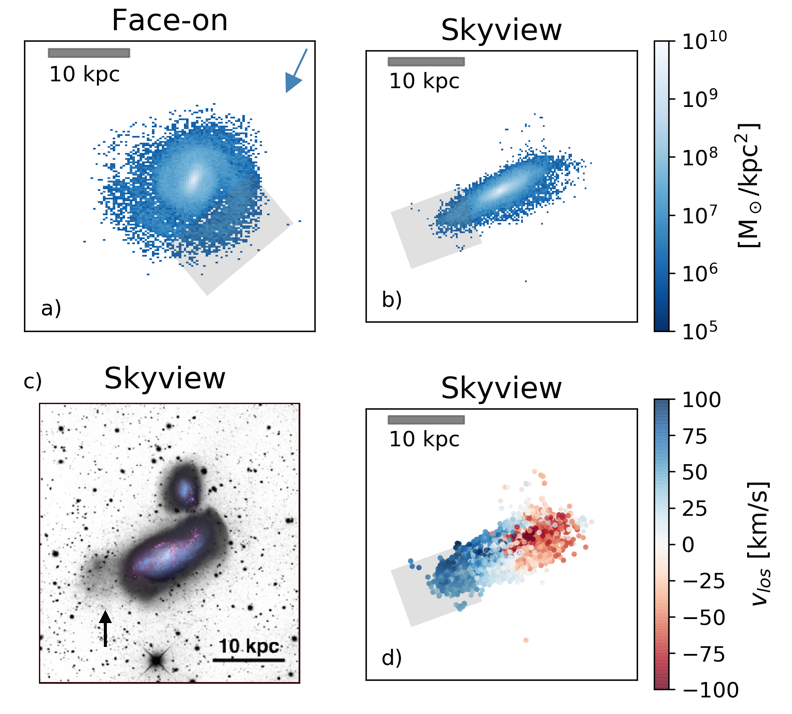}}
\caption{Projections of the primary's baryonic particles at the time of match. {\bf Panel a)} Face-on projection of the density distribution in the primary dwarf (NGC 4490) at the time of match. The blue arrow indicates our viewing direction of the system. The gray box highlights the particles in the one armed spiral (panel {\bf a}) associated with the extension of diffuse star light seen in the optical data (black arrow, panel {\bf c}). These particles are located in a one armed spiral induced by the most recent pericentric passage with the secondary (NGC 4485) where the impact parameter was $r_{p,2} = 1.7$ kpc.  {\bf Panel b)} Skyview projection of the density distribution in the primary (NGC 4490) at the time of match.  {\bf Panel c)}  Optical data of the dwarf pair (see Figure \ref{fig:n4490}). {\bf Panel d)}  Skyview of all primary baryonic particles at the time of match. The color bar shows the line-of-sight velocity ($v_{los}$) of the particles ranging from $-100$ to $100$ km s$^{-1}$. The particles associated with the extended diffuse starlight (see black arrow, panel {\bf c}) are highlighted in the gray boxes and are blue-shifted.
The recent encounter with the secondary dwarf appears to have induced a one-arm spiral mimicking the optical extension of the main body when viewed from our perspective. }
\label{fig:delgado}
\end{figure*}

In this section we investigate the effect of the repeated interactions with the secondary on the primary dwarf in our simulation. In Figure \ref{fig:moutside}, we found that the primary gets more extended and that a small fraction of its baryons reside beyond 7$r_{s,prim}$. Additionally, we presented diffuse starlight extending towards the left of the primary's main body in optical observations  in Figure \ref{fig:n4490} (right) but we did not include this feature as one of the \identikit~ matching parameters. Can this structure be caused by perturbations of the primary's stellar disk from interactions with the secondary?

In Figure \ref{fig:match}, panel {\bf a)} there is an off-centered extension of the primary disk body to the left (cyan particles).  
This is a common feature of each simulation where the secondary has a close encounter with the primary. Figure \ref{fig:match}, panel {\bf d)} shows the ``top-down" (RA vs line-of-sight distance) view of the simulation, where a spiral feature is evident in the primary (cyan) galaxy. 
In order to investigate how this structure compares to the diffuse extension of NGC 4490 seen in the optical data (Figure \ref{fig:n4490}, right), we explore the structure of the primary galaxy at the time of match in Figure \ref{fig:delgado}. In panels {\bf a} and {\bf b} we plot the face-on and plane-of-the-sky density projections of the baryons in the primary dwarf. 

The gray box is overlayed to highlight the location of the diffuse starlight in the optical observations (Figure \ref{fig:delgado}, panel {\bf c}, black arrow). The primary dwarf indeed appears to be extended in the direction of the diffuse starlight from our viewing direction (panel {\bf b}), and when viewing the primary's baryonic disk face on, we see that the extension is due to the presence of a one-armed spiral extending from the main body (panel {\bf a}). The feature is moving towards us at the time of match (panel {\bf d}). We postulate that the diffuse starlight in the plume-like feature seen in panel  {\bf c}, is extended debris from the primary. This hypothesis can be tested with future observations of the color and magnitude of the plume feature as compared to NGC 4490's main body. 

This asymmetric one-armed spiral is formed during the second pericentric passage and persists for $\sim $300 Myr (see also \citealt{lang14}, \citealt{besla16}) after which it is destroyed due to the third pericentric close passage ($r_{p,3} = 0.95$ kpc, see Figure \ref{fig:orbit}). 

The one-armed spiral appears to be a generic outcome of a collision (low impact parameter encounter) between a low mass perturber and a barred galaxy (see \citealt{Athanassoula96}, \citealt{pardy16}, \citealt{beren03}, \citealt{bekki09}, \citealt{besla12}, \citealt{besla16}). The exact location of the one-armed spiral feature is sensitive to the specific bar phase at the time of encounter, hence the details of the observed one-armed spiral might differ from the simulation result presented here.  Our simulation of the NGC 4490/4485 provides
further evidence that these types of dynamical encounters could explain the classical morphological signpost of Magellanic Irregulars (\citealt{val72}). 

At the time of match, the extension of the primary has a surface density of $\sim 10^7$ M$_{\odot}$ kpc$^{-2}$. While the exact surface density is dependent on our mass model for the primary galaxy, this value is at least a factor of 100 higher than the surface density in the envelope surrounding the pair (see Figure \ref{fig:privon13}, middle left), which can explain why we have not yet found a stellar envelope associated with the 50 kpc HI envelope. Additionally, the stellar disk scale length might be smaller than the HI disk scale length (\citealt{swaters02}), which could also limit the amount of stars in the envelope.

\section{The fate of the envelope}\label{sec:disc}
Throughout the paper, we have focused on the dynamical match to the system and what happens up until the time of match. In this section we discuss the future evolution of the system and the fate of the extended baryonic envelope. 
\subsection{Energetic and Morphological Evolution of the Envelope}\label{sec:bound}
Based on the morphological evolution of the system seen in Figure \ref{fig:privon13} it is clear that the large scale structure of the system persists and continues to evolve long after the time of match ($t = 0$ Gyr) and long after coalescence of the two dwarf galaxies ($t = 0.37$ Gyr). But will all of the envelope be re-accreted or is some unbound? And how long will it take the bound portion to re-accrete? With a dynamical match to the system, we can begin to address these otherwise observationally challenging questions.

\begin{figure}
\centerline{\includegraphics[width=\columnwidth]{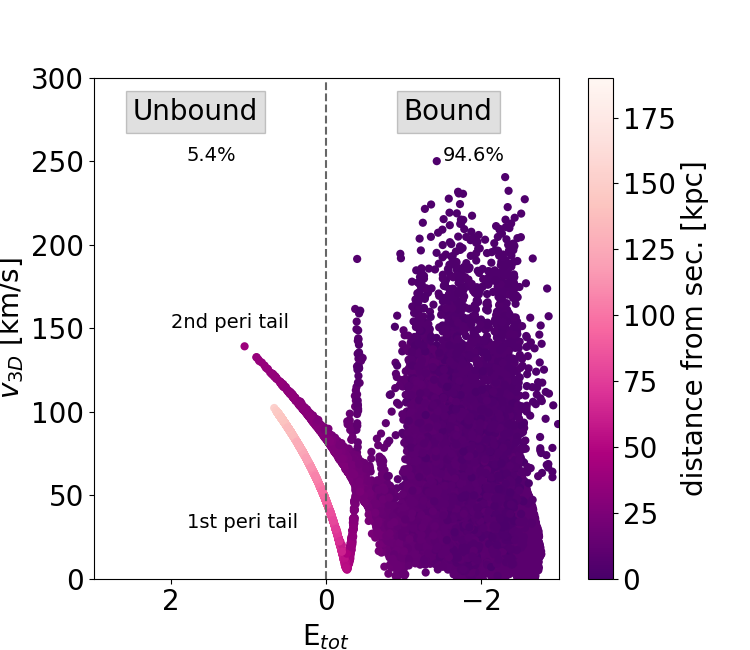}}
\caption{The magnitude of the 3D velocity vector as a function of total energy for all the secondary dwarf's baryonic particles at the time of match. The color bar indicates the 3D distance from each particle to the center of the secondary at the time of match (same color bar as in Figure \ref{fig:tailevol}).
Of the 32768 baryonic particles in the secondary dwarf galaxy, 1775 are unbound (5.4\%). The two distinct kinematic features are produced in the first and secondary pericentric passage, respectively, where the second pericentric tail particles are closer to the secondary at the time of match. Note how the tail produced in the first pass has a velocity reversal where $v_{\rm 3D} = 0$ km s$^{-1}$ close to $E_{\rm tot} = 0$. This indicates that some material in the tidal tail from the first pericentric passage is moving away from the galaxies while some material has reached its turnaround point ($v_{\rm 3D} = 0$ km s$^{-1}$ relative to the center) and has started to fall back. }
\label{fig:bound}
\end{figure}

In Figure \ref{fig:bound} we plot the 3D velocity, $v_{\rm 3D}$, of each secondary baryonic particle in the simulation (32768 particles) as a function of their total energy, $E_{\rm tot}$, at the time of match. The color bar shows the 3D distance of each particle from the center of the secondary dwarf. The total energy is calculated as:
\begin{equation}
E_{\rm tot} = \frac{1}{2} (v_{\rm 3D} - v_{\rm prim})^2 + \phi,
\end{equation}  
where $v_{prim}$ is the magnitude of the center of mass velocity of the primary at the time of match and $\phi$ is the potential energy of each particle based on their location, stored as an output for each particle in each snapshot of the simulation. If the particles have a negative total energy they are bound to the center of mass of the combined system of the primary and the secondary galaxy. 

The tidal features produced in the first two pericentric passages separate as two distinct kinematic features. The part of the tail produced in the first pericentric encounter ($t = -1.29$ Gyr, see Figure \ref{fig:orbit}, left) is located farthest away ($d_{3\rm D} \sim 175$ kpc), while the tidal debris stripped in the more recent second pericentric passage ($t = -0.23$ Gyr, see Figure \ref{fig:orbit}, left) has only reached $d_{3\rm D}$  $\sim 34$ kpc at the time of match. 

At the time of match only 5.4\% of the total baryonic mass of the secondary is unbound and is removed at either the first or second pericentric passage. From Figure \ref{fig:moutside} we know that $\sim$ 34\% of the secondary particles reside beyond 7 disk scale radii at the time of match and that the rest of the material is within the main body at small distances (see dark purple points). If we only include the particles beyond 7 disk scale radii of the secondary (7 $\times~ r_{s,sec}$ = 2.1 kpc) at the time of match in our calculation, 18.9\% of those secondary particles are unbound. 

Figure \ref{fig:bound} shows that the tail from the first pericentric passage exhibits a velocity reversal (where part of the bound tidal tail has $v_{3\rm D}=0~ $km s$^{-1}$), indicating that some material in the tidal tail from the first pericentric passage is moving away from the galaxies while some material has reached its turnaround point ($v_{3\rm D}=0~ $km s$^{-1}$) and has started to fall back.

\begin{figure*}
\centerline{\includegraphics[width=\textwidth]{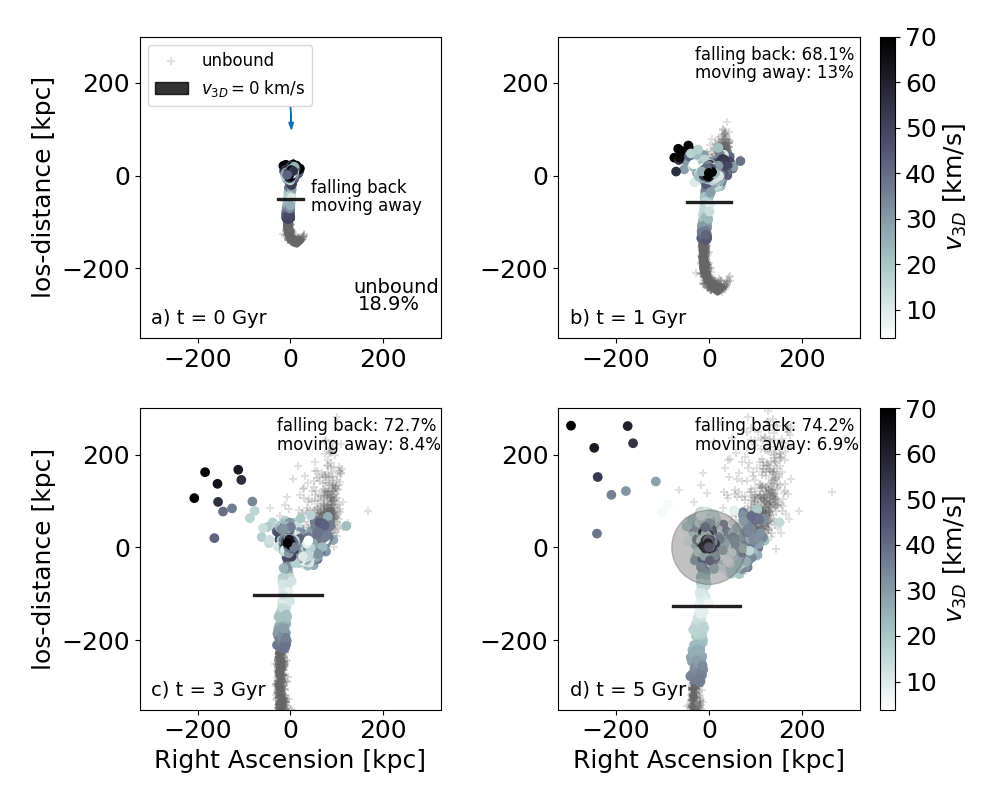}}
\caption{The ``top-down" (RA vs los-distance) morphological evolution in time (panel {\bf a} through {\bf d}) of the secondary dwarf galaxy's particles that were beyond 7 disk scale radii at the time of match. The color bar denotes the 3D velocity (y-axis from Figure \ref{fig:bound}). We fix the maximum value of the color bar to $v_{3\rm D} = 70$ km s$^{-1}$. The gray "+"s show the 18.9\% of particles beyond 7 disk scale radii at the time of match that are unbound and which will continue to move away from the system. This number will remain the same between all panels, and we do not include these particles in our analysis of material ``moving away" and ``falling back" to the merger remnant. The blue arrow in panel {\bf a} indicates our viewing perspective of the system. The black bars demonstrate the turnaround-radius (apocenter/zero-point-velocity) at which the particles in the tidal tail from the first pericentric passage shift from moving outwards to start falling back towards the center of mass (white particles have $v_{3D} = 0$ km s$^{-1}$). As time passes, more and more particles start to fall back towards the center of mass (see percentages) and the  $v_{3D} = 0$ km s$^{-1}$ turnaround points move further out along the tidal tails to larger distances. The gray circle in panel {\bf d} illustrates an 80 kpc sphere, which encloses the particles that have fallen back from the tidal tails and orbit the center of mass. After 5 Gyr (panel {\bf d}) much of the debris remains in the two tails produced during first and second pericentric passages, demonstrating that the large scale structure persist for several Gyr after the dwarfs coalesce. 
}
\label{fig:bound_track}
\end{figure*}

Motivated by this velocity reversal, we investigate the future evolution of the secondary's debris beyond 7 disk scale radii at the time of match.  
In Figure \ref{fig:bound_track}, we show 4 ``top-down" (RA vs los-distance) simulation snapshots tracking the future fate of the tidally stripped material at the time of match. The unbound particles (18.9\%) are marked as ``+" symbols. 

In Figure \ref{fig:bound_track}, we do not explicitly track material tidally removed from within 7 disk scale lengths in future passages, as we are only tracking the present day envelope. We expect that the later tidal tails are generally much less extended and should re-accrete more rapidly.
The blue arrow indicates our viewing direction towards the system at the time of match. The color bar denotes the 3D velocity (y-axis from Figure \ref{fig:bound}). We fix the maximum value of the color bar to $v_{3\rm D} = 70$ km s$^{-1}$ such that the color bar is not dominated by the high velocity particles within the secondary's disk (see y-axis from Figure \ref{fig:bound}). This enables us to better illustrate the velocity gradient along the tail stripped at the first pass (some material is starting to fall back, white particles: $v_{3D} = 0$ km s$^{-1}$). The black bars in Figure \ref{fig:bound_track} indicate where $E_{\rm kin} = v_{3\rm D} = 0$ for the tail produced in the first pericentric passage. All tidal material in the tail closer to the center of mass than this limiting distance (black bars) is falling back and returning to the primary. 

Panel {\bf a)} of Figure \ref{fig:bound_track} represents the current state of the system (the time of match). At this point in time the secondary has undergone two pericentric passages, creating two kinematically distinct tidal tails (see Figure \ref{fig:bound}). In this projection of the simulation (``top-down"), the particles stripped in the second pericentric passage, are moving mostly in the negative $z$(positive declination)-direction. Therefore, we do not plot the velocity reversal black bars for the tail produced in the secondary pass, however we do account for these particles in our mass budget of material moving away vs returning to the system. 

Panel {\bf b)} represents the system 1 Gyr into the future. The debris from the second pericentric passage tail is growing in length (to more negative $z$-values in this projection) and some of the material starts to fall back towards the center of mass.  The secondary has made a third passage (see Figure \ref{fig:orbit}), generating more tidal debris (towards more negative RA values). Recall that we are only tracking debris which was already beyond 7 disk scale radii at the time of match (in principle more debris from within 7 disk scale radii will also be tidally moved in this third pass).  At this point in time 68.1\% of  the bound present day envelope is returning to the merger remnant, 13\% is moving away from the remnant (some of this might be unbound in the third pass) while 18.9\% of the particles which were already unbound at the time of match are still moving away (gray ``+" markers). 

In Panel {\bf c)}  it is evident that the location of the velocity reversal (black bar) has moved even farther out along the first tidally produced tail (to more negative los-distance values). This is due to the fact that the particles along the tail lose kinetic energy as time passes and start falling back towards the center of mass (see percentages).

Panel {\bf d)} represents the system 5 Gyr into the future and we have overlayed an 80 kpc sphere, which roughly represents the turnaround radius for debris that has returned from the tails and has started to orbit the center of mass of the merger remnant.  This limiting radius of 80 kpc sphere was determined by investigating particle motion of the returned debris at $t = 5$ Gyr. We seek to track the motion of particles across this radius to infer the rate at which the present day tidal debris will return to the system.  In this work, we are investigating a collisionless $N$-body simulation (i.e. without hydrodynamical effects), but we would not expect the gas in the remnant to extend beyond the collisionless particles (i.e. beyond 80 kpc), as the gas should dissipate energy likely resulting in a more compact configuration than we see here. See Section \ref{sec:hydro} for a discussion of hydrodynamical effects. 

Interestingly, after 5 Gyr (panel {\bf d}) much of the debris remains in the tails produced during the various pericentric passages, demonstrating that the large scale structure may persist for several Gyr after the dwarfs coalesce, and that the baryons can be ``parked" at large distances for long timescales. Moving material to large distances in a galaxy encounter is not unique to dwarf galaxy interactions. However, as dwarfs typically have larger gas to stellar fractions than more massive galaxies, a substantial fraction of potential future fuel for star formation is likely ``parked" at large distances for dwarf encounters. A comparative study of gas re-accretion post dwarf mergers and massive galaxy mergers has yet to be done. It is important to note that the timescales involved will be affected by our assumed galaxy mass model (see Section \ref{sec:rate}). For a discussion on return of tidal material in a $\mu = 1:1$ mass ratio galaxy merger, see  \citet{hibbard95}.

\subsection{Rate of return of the envelope}\label{sec:rate}
\begin{figure*}
\centerline{\includegraphics[width=\textwidth]{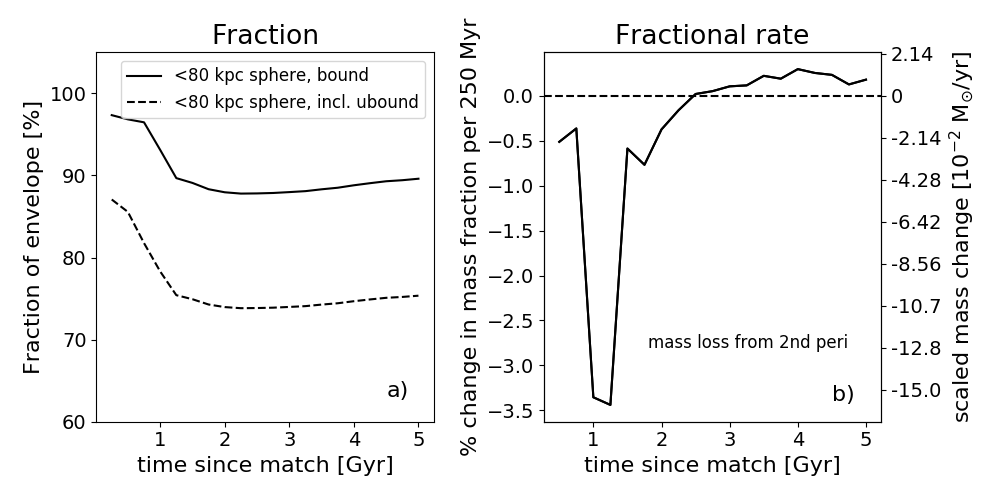}}
\caption{{\bf Panel a)} Fraction of the total baryonic envelope that resides within a sphere of 80 kpc in radius centered on the primary (NGC 4490) (illustrated in Figure  \ref{fig:bound_track} panel {\bf d}) as a function of time after match (in the future). The solid line shows the fraction using only the bound particles in the envelope at the time of match. The dashed line shows the same calculation, but now including the unbound particles (see "+" markers in Figure \ref{fig:bound_track}). {\bf Panel b)} Fractional rate of mass moving in/out of the 80 kpc sphere for the bound particles as a function of time after match. Left y-axis demonstrates the percentage of all the bound particles in the simulation that are moving out of/into the 80 kpc sphere. The right y-axis shows this same rate converted to M$_{\odot}$/Myr and scaled based on the mass in the present day observed envelope (M$_{\rm HI}$(outside)$ = 1.07 \times 10^9 ~\msun$). The dashed horizontal line shows where no mass accretion nor mass loss occurs. The large drop in the fractional rate occurs as the material from the secondary pericentric pass moves out of the 80 kpc sphere. From both panels it is evident that mass continues to be lost for several Gyr after the time of match, but that after $>$ 2 Gyr the re-accretion of debris from both the first and second pericentric dominates over the fraction of mass flowing out of the 80 kpc sphere.}
\label{fig:rate}
\end{figure*}

It is very difficult to assess the inflow rates of baryons to galaxies observationally and only few examples exist (e.g. \citealt{zheng17}).
However, when simulating a dynamical encounter, we can estimate the inflow rate of baryons to the merger remnant (see also \citealt{hibbard95}). 
We use the 80 kpc radius radius sphere described in Section \ref{sec:bound} and shown in Figure \ref{fig:bound_track} panel {\bf d}, to track the future evolution of all material in the present day envelope. We define the envelope to be all material beyond 7  $\times~ r_{s,sec}$ (2.1 kpc) at the time of match. We can thus track the inflow of material from the outer HI envelope in time.

In Figure \ref{fig:rate} we track the tidal debris that is in the envelope at the time of match. In particular, we investigate fractionally how much of this material is residing within vs outside the 80 kpc sphere as a function of time (Figure \ref{fig:rate}, panel {\bf a}). Here, we distinguish between the bound (solid line) and total (bound+unbound; dashed line) debris. As the unbound material will not return to the system, it will not affect the mass inflow to the system (or rate of return).

Up until $\sim$ 2 Gyr after the time of match, the 80 kpc sphere loses more mass than it gains, hence the bulk of the tidal material is still moving away from the system, which is consistent with the fact that the second pericentric tail is leaving the system during this time (see Figure \ref{fig:bound_track}). As time progresses more material from the outer envelope returns to reside within the 80 kpc sphere. After 2.5 Gyr we see a net inflow of the outer envelope to the 80 kpc sphere, hence more material is flowing into than out of the sphere. This implies that the material within this radius (i.e. the bulk of the envelope) will return to the merged system.  
The total  bound mass in the simulated envelope shown in Figure  \ref{fig:rate}, panel {\bf a}, corresponds to $\sim 1.4 \times 10^8$ M$_{\odot}$ and after 5 Gyr $\sim 10$\% of the envelope has yet to be re-accreted. 

This behavior is more clearly seen in Figure \ref{fig:rate}, panel {\bf b}, where we plot the fractional rate of return of bound material within the 80 kpc sphere. The second pericentric passage causes a net outflow of material, but after 3 Gyr there is a net inflow.  Scaling the fractional values using the total mass of the observed envelope implies a net inflow rate of 0.01 \msun~ yr$^{-1}$ at 80 kpc (see right y-axis).

We have utilized light dark matter halos in this study. \citet{barnes16} showed that the ratio of the escape velocity ($v_e$) from a galaxy and its circular disk velocity ($v_c$) at the half mass radius ($\mathcal{E} = \frac{v_e^2}{v_c^2}$) can greatly affect the radii to which tidal material reach after tidal removal and its re-accretion rate. If $\mathcal{E}$ is higher the galaxies reabsorb their tails faster. Using more massive halos, as expected from abundance matching, would result in the debris reaching smaller radii before velocity reversal and that the acceleration is higher as the debris falls back due to the deeper potential well. The results presented in this section should therefore be interpreted as lower limits on the inflow rate of the outer envelope. Given that the bulk of the envelope remains within 80 kpc of the system even in our low mass halo models, we conclude that the observed envelope will provide a long lived gas supply channel to the merged remnant. 

\section{Discussion}\label{sec:disc2}
In this section we compare our results to previous simulations of the Magellanic System (Section \ref{sec:lmcsmc}), and we discuss the implications of our work in the context of tidal pre-processing and ongoing dwarf galaxy surveys (Section \ref{sec:tnt}). Finally, we discuss baryon cycles in dwarf galaxies in the context of gas inflows and dissipational effects (Section \ref{sec:hydro}). 

\subsection{Comparison to the LMC \& SMC system}\label{sec:lmcsmc}
\citet{besla12} showed that the mutual interaction between the Magellanic Clouds (MCs) can reproduce both the large-scale gaseous distribution (bridge, leading and trailing arm) of the Magellanic System along with internal properties of the clouds, irrespective of tides from the Milky Way. 
Our match to the NGC 4490/4485 system is similar to the LMC/SMC models prior to their infall to the Milky Way (see \citealt{besla16}) as both solutions 
suggest that a substantial amount of baryons are moved to large distances from the secondary galaxy (i.e. NGC 4485 and the SMC) in a prograde encounter. 

Additionally, the NGC 4490/4485 system might provide clues to what the LMC/SMC would have evolved into if they had not been in the vicinity of the Milky Way (see \citealt{besla16}). Because the LMC/SMC are passing close by the Milky Way, the SMC might be unbound from the LMC (\citealt{gonzalez16}), which has been suggested based on proper motion measurements of the clouds (\citealt{kalli06}, \citealt{vieira10} \citealt{kalli13}, \citealt{zivik18}, \citealt{niederhofer18}). In the \citet{besla16} simulation (LMC/SMC evolving without the Milky Way), the Magellanic Clouds' present day 3D separation is 26 kpc ($t \sim$ 6.3 Gyr in their figure 5), and the most recent orbit has a timescale of $\sim$ 800 Myr. For comparison the 3D separation between NGC 4490/4485 in our best match is only 9.3 kpc at present day and the orbital timescale of the last orbit is only $\sim$330 Myr (see Figure \ref{fig:orbit}, left). This is similar to what the LMC/SMC orbit would yield after one more pericentric passage close to the next apocenter (\citealt{besla16}, figure 5), suggesting that the NGC 4490/4485 pair is a more evolved version of the Magellanic Clouds, had they never fallen in to the Milky Way. Both our orbit and the \citet{besla16} orbit have high eccentricities at the time of their best match ($e=0.83$ and $e=0.66$, respectively), but the two orbits are not exactly analogous.
Our models thus expand the range of plausible orbital parameters for dwarf-dwarf binary galaxy encounters. Both the LMC/SMC and NGC 4490/4485 solutions predict that in isolation, both systems would rapidly merge without the intervention of a massive third body. Additionally, our work demonstrates that it is indeed possible to have a large extended gases envelope form through tides alone, providing new observational and theoretical support for the scenario presented for the LMC/SMC in \citet{besla10}, \citet{besla12}, \citet{besla16}.

A recent example of a dwarf pair that also mimics the LMC/SMC system was reported by \citet{paudel17}. The pair (UGC 4703/4703B) is very gas rich, shows signs of interaction (e.g. HI bridge, SF in bridge, extended gas) and the individual galaxies reside 81 and 104 kpc from their MW-type host, respectively. This pair provides an interesting example in addition to NGC 4490/4485 for understanding the effect of a mutual dwarf-dwarf tidal interaction, prior to infall to a massive host (i.e. pre-processing). 
In future work, we plan to assess the statistics of orbital configurations for dwarf encounters to advanced our understanding of the dwarf-dwarf merger sequence. See also \citet{besla18} for a theoretical comparison between the frequency of dwarf-dwarf encounters in \textit{Illustris} and SDSS at low redshift.

\subsection{Tidal pre-processing}\label{sec:tnt}
The fact that baryons can remain extended for several Gyr after coalescence of interacting dwarf galaxies is important for understanding how tidal pre-processing between dwarfs can affect gas removal from low mass galaxies (quenching) after infall to more massive/gas rich environments. Ram-pressure and tides from a nearby massive galaxy seem to be inefficient at stripping gas that is tightly bound to the dwarfs (e.g. \citealt{emerick16}, \citealt{fillingham16}). If the gas is much more extended, removing it from the dwarfs through e.g. tides from a host galaxy or ram-pressure stripping by a hot halo of a host galaxy will be more efficient (e.g. \citealt{emerick16}, \citealt{salem15}) and extended gas structures caused by dwarf-dwarf interactions could help this process (\citealt{pearson16}). 

This picture appears consistent with the $\Lambda$CDM theory as \citet{wetzel15} have shown that 30-60\% of dwarf satellites are expected to have been accreted as part of a low mass group. Furthermore, \citet{marasco16} found that quenched dwarf satellites of MW type hosts at $z = 0$ preferentially experienced a satellite-satellite encounter prior to accretion. 

From the TiNy Titans Survey, we know that dwarf galaxy pairs are just as gas rich as non-paired dwarfs if they reside far from a massive host galaxy (\citealt{stierwalt15}), but that much of this gas can be located in extended tidal features (\citealt{pearson16}). The work presented in this paper demonstrates that in encounters where gas is moved to large radii, gas may remain extended for several Gyr. These long timescales indicate that we should expect to find gas in large extended structures surrounding the dwarfs when surveying dwarf galaxy pairs (e.g. \citealt{stierwalt15}), dwarf groups (e.g. \citealt{stierwalt17}) and dwarfs with merger driven starbursts (e.g. \citealt{lelli12}). Interestingly, there is a pair of dwarf galaxies (NGC 4618 \& NGC 4625: \citealt{pearson16}) with stellar masses of $M_* = 4.3 \times 10^{9} ~\msun$ and $M_* = 1.3 \times 10^{9} ~\msun$, respectively in the vicinity of NGC 4490/4485.  NGC 4618/4625 pair has a velocity separation of only $\Delta v = 20 ~\kms, $\footnote{From NED redshifts: $\Delta v = c ~\times$ \textbar $z1 - z2 $\textbar$ / (1 + (z1 + z2)/2)$.} with respect to the NGC 4490/4485 pair, and they are at a projected separation of $d_{\rm proj} \sim$ 267 kpc, assuming a distance of 7.14 Mpc (as for NGC 4490/4485). The NGC 4618/4625 pair is at a larger separation than reported for the seven newly discovered dwarf groups in \citet{stierwalt17} (their projected separations are $<80$ kpc), and it is unlikely the NGC 4618/4625 pair has had a dynamical influence on the NGC 4490/4485 pair due to the large projected separation. \citet{besla18} shows that cosmologically it should be rare to find groups of dwarfs at low redshift (0.013 $<$ z $<$ 0.0252) with stellar masses of the members larger than $M_* > 2 \times 10^8 ~\msun$, although they use a more conservative search criterion of  $15 < d_{\rm proj} < 150$ kpc and $\Delta v < 150 ~\kms$.

The COS-Dwarfs Survey (\citealt{bordiloi14}) find substantial amounts of ionized gas at large distances ($\sim$ 110 kpc) surrounding 43 dwarf galaxies at low redshift ($z< 0.1$) and they suggest a wind-driven origin of the gas. While it remains unclear whether the tidally removed gas in our dwarf-dwarf encounter will be and/or remain ionized as a result of the interaction (see \citet{weil17} for an example and discussion of ionized gas in a massive galaxy merger), our work presents another contributing factor to gas at large distances as we find baryons orbiting the dwarf merger remnant long after coalescence (see Figure \ref{fig:privon13} and \ref{fig:bound_track}).

\subsection{The Baryon Cycle and Unmodeled Hydrodynamic Effects}\label{sec:hydro}
The current star formation rate of NGC 4490 inferred from the FUV non-ionizing continuum is SFR$_{\rm N4490}$(FUV) = 1.9 \msun~ yr$^{-1}$ (\citealt{lee09}, scaled to a distance of 7.14 Mpc here). We showed that $\sim$90\% of the gas envelope is within a sphere of 80 kpc after 5 Gyr and that the material beyond the 80 kpc sphere (the outer envelope) is flowing into the region at a roughly constant rate of $\sim 10^{-2}$ M$_{\odot}$ yr$^{-1}$ after 3 Gyr, representing a lower bound on the inflow rate. We expect all of the bound material (within and beyond the 80 kpc sphere) to accrete back on to the merger remnant at some point, potentially refueling star formation. However, to quantify the exact properties of the accretion rate of material onto the merger remnant itself, we need hydrodynamical simulations.
 
In this work, we have narrowed down the parameter space of the tidal interaction between the two dwarf galaxies NGC 4490 and NGC 4485. Our work is based on collisionless $N$-body simulations and we do not expect our global match to be much affected by dissipational effects as the large scale features  (e.g. the symmetric HI envelope) and orbital decay are not strongly affected by the inclusion of gas (\citealt{barnes96}). 
However, to properly study the structure of the remnant and where the accreted gas goes, hydrodynamical effects will be important to include in future simulations.

Additionally, we need hydrodynamical simulations to investigate the HI bridge connecting NGC 4490/4485 (see Figure \ref{fig:n4490}, left). Based on our collisionless particle simulation presented in Figure \ref{fig:match}, we do not see evidence of a tidal bridge at the location of the dense bridge material (see red arrow in Figure \ref{fig:match}), which indicates that this is a hydrodynamical feature. In our match, the orbit of the secondary passes through the dense gas bridge observed in HI during the first two pericentric passages in Figure \ref{fig:match}, panel {\bf a} and in the first pericentric passage in Figure \ref{fig:match}, panel {\bf b}, {\bf c}.  Hence a hydrodynamical origin of the bridge seems plausible as ram-pressure effects from a collision between two gaseous disks can increase the amount of gas lost to the bridge region (see the "Taffy bridge" for an example of this effect: \citealt{condon93}, \citealt{gao03}).

\section{Conclusion}\label{sec:con}
In this paper, we have computed the first dynamical match to an observed isolated galaxy encounter at the low mass, dwarf scale using \identikit~ with $N$-body follow-up. The system NGC 4490/4485 is an analog of the Magellanic System prior to its infall to the Milky Way, surrounded by a massive HI envelope but located in isolation from any massive galaxy. Our results and conclusions are summarized as follows:

\begin{itemize}
\item[1.]{We are able to find a kinematic and morphological match to the gas distribution of the dwarf pair NGC 4490/4485 through solely the tidal forces from the primary (NGC 4490) stripping material from the secondary (NGC 4485). The match to the pair required a solution in which the secondary dwarf galaxy's  (NGC 4485's) spin is prograde to the orbit and has a high inclination orbit with
respect to the more massive dwarf (NGC 4490).} 

\item[2.]{In this match the $\sim$50 kpc (projected) envelope consists of a large tail from NGC 4485, lost during the first pericentric encounter between the two galaxies ($\sim$ 1.4 Gyr ago). Due to our viewing perspective we see the tail as an envelope wrapping around the entire system. This demonstrates that
through tidal interactions between two low mass galaxies, gas can be moved to large distances and produce a massive, symmetric, neutral HI envelope, without the need for stellar feedback (see \citealt{clemens98})} or perturbations from a massive host.

\item[3.]{We predict that NGC 4490 \& 4485 will fully merge in 370 Myr, but that the gaseous envelope will remain extended well after coalescence.}\\

\item[4.]{During the encounter a one-armed spiral is induced in primary dwarf (NGC 4490) which is also seen in the new optical data of the system presented in this paper. This demonstrates that a high mass ratio, small impact parameter collision between two low mass galaxies can explain the classical morphological signpost of Magellanic Irregulars (\citealt{val72}).}\\

\item[5.]{The fact that the extended tidal features evolve and persist for several Gyr after coalescence, supports the idea that dwarf-dwarf interactions play an important role aiding gas removal and quenching of low mass galaxies. Subsequent ram-pressure stripping and even weak tidal forces will be much more efficient to remove gas from such extended structures formed by pre-processing.}

\item[6.]{We studied the current dynamical state of the HI envelope around the NGC 4490/4485 system and concluded that baryons are ``parked" and eventually re-accreted by the merger remnant over long timescales ($>$ 5 Gyr). The material in the extended tidal features will return to the system, crossing an 80 kpc radius at a rate of 0.01 \msun~ yr$^{-1}$ within 2.5 Gyr. The bulk of the HI envelope will be re-accreted to the merged remnant, providing a long-lived supply channel of gas. If dwarfs in the field with large gas reservoirs (e.g. \citealt{muerer96}, \citealt{werk10}, \citealt{kreckel11}) have had a previous merger, the long accretion time can help explain the limited star formation in their gaseous outskirts.}

\item[7.]{This work provides a novel isolated analog to simulations of the LMC/SMC. We illustrate that, generically, a significant fraction of the gas can be moved to large distances in prograde dwarf-dwarf interactions, without the aid of a massive host (see also \citealt{besla12}, \citealt{dongia10}). Additionally, we find that the NGC 4490/4485 orbit appears to be an evolved version of what the LMC/SMC might have looked like had they not been accreted by the Milky Way.}
\end{itemize}

\acknowledgements
We thank Sabrina Stierwalt, Kelsey E. Johnson, Sandra E. Liss, Joshua Barnes, Kelly Blumenthal and Andrew Emerick for insightful discussions. We thank Marcel Clemens for sharing his data of NGC 4490 \& NGC 4485. The Zeno software package was used in this research. The Numpy software package (\citealt{walt11}) and Matplotlib graphics package (\citealt{hunter07}) were used in this research. This research has made use of the NASA/IPAC Extragalactic Database (NED) which is operated by the Jet Propulsion Laboratory, California Institute of Technology, under contract with the National Aeronautics and Space Administration. This material is based upon work supported by the National Science Foundation under Grant No. NSF grant 1714979 and NSF grant 1715944. We also acknowledge support from NASA for program 13383 from the Space Telescope Science Institute, which is operated by the Association of Universities for Research in Astronomy, Inc., under NASA contract NAS5-26555. KVJ's contributions were supported by NSF grant AST-1614743. G.C.P. acknowledges support from the University of Florida. NK is supported by the NSF CAREER award 1455260. A portion of this work was performed at the Aspen Center for Physics, which is supported by National Science Foundation grant PHY-1607611. DMD acknowledges support by Sonderforschungsbereich (SFB) 881 ``The Milky Way System'' of the German Research Foundation (DFG), particularly through sub-project A2. This research has made use of the NASA/IPAC Extragalactic Database (NED), which is operated by the Jet Propulsion Laboratory, California Institute of Technology, under contract with the National Aeronautics and Space Administration.

\appendix\label{sec:appendix}
Throughout the paper, we have presented a kinematic and morphological match to the low mass, isolated galaxy encounter NGC 4490/85, which reproduces key features of the HI and optical data. In this appendix we discuss some of the alternative configurations which could reproduce some, but not as many of the features of the encounter. \\

\begin{itemize}
\item{\bf Number of pericentric passes}\\
A natural starting point when searching for a match is ensuring that the positions of the galaxies end up at the correct locations in morphology and velocity space. A natural next step is to explore how many pericentric passages are needed. The match presented in the paper occurs between the second and third pericentric pass close to apocenter. When exploring other matches, we found that any scenario in which only one pericentric passage had occurred did not provide a good match, as the tail from the first passage had not grown to be large enough to populate the envelope if the sizes of the galaxies (and separations) had to match the data.  This was the case even in the most prograde scenario with substantial mass loss from the galaxies. 

When moving to a third pericentric pass, we could reproduce the overall morphology and kinematics of the encounter, often with a slightly larger length scaling (to fit the extent of NGC 4490 HI disk), however the tidal tail from the secondary galaxy produced in the first passage grew to be much too long and extended beyond the extent of the HI envelope in the wrong direction. 

\item{\bf Wider passes}\\
We also explored the initial separation. A closer initial pass will lead to a more dramatic mass loss (depending on the disk angles). The match presented in this paper has an idealized pericentric separation at first pass of $4.2 \times r_{disk,prim}$. We explored wider pericentric separations for the encounter, but we were not able to populate the south of the HI emission kinematically (see Figure \ref{fig:match}, panel {\bf b}). In the wider passes the angle of the tail produced in the first pass became too wide and did not populate the south of the envelope with the tail ``wrapping around" from the first pass. We tested that moving to a later interaction stage and a different viewing angle did not resolve this problem. 

The widest pass we explored was a pass with an initial pericentric separation ($r_{peri} \sim 20~ r_{disk,prim}$). Between the third and fourth pass, we could get a match similar to the one presented in this paper. However, some debris from the first pass was present and predicted emission where none was observed. Hence, we did not obtain a better match by adding another pass. 

\item{\bf Populating north and south from the secondary and primary, respectively}\\
We explored encounters in which the north of the envelope morphologically and kinematically was solely populated by debris coming off of the secondary galaxy in the second pericentric pass, and in which the south was populated by debris from the primary galaxy kinematically and morphologically. Given the known high stellar mass ratio between the two galaxies, we were not able to populate the HI envelope symmetrically morphologically (the south was unpopulated) nor populate the kinematic features with this approach. In particular the emission located where the white arrows point to in Figure \ref{fig:match} was left unpopulated. The observed baryonic mass ratio of the pair is $\mu \sim 8:1$, however it is possible that the dark matter halo mass ratio is lower (see Section \ref{sec:prop}). Going to more equal mass ratios would probably violate the observed stellar mass ratio.

\item{\bf Populating the ``bridge region"}\\
As discussed in Section \ref{sec:hydro}, the orbit of the match presented in the paper passes through the dense bridge region in Figure \ref{fig:match},  panel {\bf b)} (see red arrow) in the first pericentric pass, and the orbit passes through the bright HI emission in the panel {\bf a)} (see also Figure \ref{fig:n4490}, left, inner HI contour) in both the first and second pass. The fact that the secondary's orbit passes through the primary's disk and through the bright HI emission associated with the bridge, provides a plausible scenario for forming a gas bridge due to ram-pressure effects (e.g. \citealt{besla12}). 

As a bridge could be a transient phenomenon, we also explored matches for which the orbit passed through the bright HI emission of panel  {\bf b)} in Figure \ref{fig:match} (red arrow) in the second pass instead of the first. Using this constraint, we could find a match for which the overall properties were similar (scaling, separations, morphology) to our presented match. However, we could not reproduce the kinematics of the tail from the first pericentric pass in this scenario (see white arrows in Figure  \ref{fig:match}, panel {\bf b/c}) and the southern part of the emission was not as populated (Figure  \ref{fig:match}, panel {\bf a}) as the tail from the first pass did not wrap around the system which is the case in our presented match. Hence, we discarded this match in favor of the match presented in the paper. 

\end{itemize}

While there is some room for changing the exact parameters of the encounter (see Section \ref{sec:match}), the match presented in the paper was the most satisfactory in terms of populating the northern and southern parts of the HI data morphologically and kinematically, reproducing the present day positions (i.e. separation) of the galaxies morphologically and kinematically, matching the observed velocity scale of the data, and having an orbit that passes through the dense bridge material in Figure  \ref{fig:match}, panel {\bf a, b, c}. 
Hence, we have presented a plausible scenario for the encounter geometry of the NGC 4490/85 system, yet further explorations of the eccentricity of the orbit, specific mass models of the galaxies and adding dissipational effects might change the specifics of the match. However, we do not expect the overall encounter geometry, which was common between all matches discussed above, to change (prograde orbit enabling substantial mass loss from the secondary galaxy and high inclination orbit such we see the tail in projection from our viewing direction).

\end{document}